\documentclass[12pt,reprint, aps, prd, showpacs,nofootinbib]{revtex4}

\usepackage[utf8]{inputenc}
\usepackage[T1]{fontenc}
\usepackage{amsmath,amssymb}
\usepackage{hyperref}
\usepackage{graphicx}
\usepackage{subfigure}

\usepackage[utf8]{inputenc}
\usepackage[T1]{fontenc}
\usepackage{amsmath,amssymb}

\begin{document}

\title{The $\bf {\it SL}(2,\mathbb{R})$ totally constrained  model: three quantization approaches.}
\author{Rodolfo Gambini}
\email{rgambini@fisica.edu.uy}
\affiliation{Instituto de F\'{i}sica, Facultad de Ciencias, Igu\'a 4225, esq.\ Mataojo, Montevideo, Uruguay}
\author{Javier Olmedo}
\email{jolmedo@fisica.edu.uy}
\affiliation{Instituto de F\'{i}sica, Facultad de Ciencias, Igu\'a 4225, esq.\ Mataojo, Montevideo, Uruguay}

\begin{abstract}
We provide a detailed comparison of the different approaches available for the
quantization of a totally constrained system with a constraint  algebra
generating the non-compact $SL(2,\mathbb{R})$ group. In particular, we consider
three schemes: the Refined Algebraic Quantization, the Master Constraint
Programme and the Uniform Discretizations approach. For the latter, we provide a
quantum description where we identify semiclassical sectors of the
kinematical Hilbert space.  We study the quantum dynamics of the
system in order to show that it is compatible
with the classical continuum evolution.  Among these quantization approaches,
the Uniform Discretizations provides the  simpler description in agreement with
the classical theory of this particular  model, and it is expected to give new
insights about the quantum dynamics of  more realistic totally constrained
models such as canonical general relativity.
\end{abstract}

\maketitle

\section{Introduction}

Fully constrained models are a class of settings that successfully describe the
physics of systems largely characterized by certain symmetries. However, the
quantization of fully constrained systems encounters several obstacles
undermining the validity of the resulting microscopical description. The usual 
strategy for first class systems, proposed originally by Dirac~\cite{dirac},
consists of representing the constraints as quantum operators on a given Hilbert
space, identifying the quantum observables and states invariant under the
symmetries generated by the constrains and, if they give rise to a large enough
set, endow them with a Hilbert space structure. One of the most prominent
examples is gravity, which turns out to be diffeomorphism invariant. In
particular, if one tries to carry out a quantum description of general
relativity, the implementation of the constraints seems necessary  if one is
interested in studying the full quantum dynamics. For instance, the lack of a  consistent implementation of the quantum constraints that reproduce in the classical limit general relativity is one of the reasons because the quantization
programme of Loop Quantum Gravity \cite{lqg1,lqg2,lqg3} is still incomplete.
This is usually known as the problem of dynamics and it is the main problem that we address in this paper for the particular case of the $SL(2,\mathbb{R})$ model.

In the last years, several approaches  have emerged attempting to shed light on
the fundamental description of this kind of models. One of the quantization
programmes that will be considered in this manuscript is the so-called Algebraic
Quantization~\cite{aq1} (and its more modern version known as Refined Algebraic
Quantization~\cite{almmt,raq}). In particular, in this approach one assumes that
given a kinematical Hilbert space $\cal H_{\rm kin}$ and a $*$-algebra $\cal
A_{\rm obs}$ of observables, the latter are represented on a dense, linear
subspace $\Phi\subset\cal H_{\rm kin}$, while the constraints have solutions
belonging to the algebraic dual $\Phi^*$ of $\Phi$. The final step consists of
endowing with a suitable Hilbert space structure this space of solutions,
following on the physical Hilbert space, and looking for a suitable
representation of Dirac observables on it. In many cases this last step is
accomplished  by applying, e.g., the group averaging technique~\cite{group-av}.
This method can be applied  without major difficulties if one knows an invariant
domain of the quantum constraints. But in occasions such a domain is not known
or pathological in so far as it either prevents the application of the group
averaging technique or  introduces ambiguities that require additional inputs in
order to achieve a consistent quantum theory. A complementary procedure  is
available when a complete set of observables is known, and requires reality
condition on them. This requirement turns out to uniquely select an inner
product~\cite{rendal}, and the completion of the space of solutions with gives
the physical Hilbert space.

A second possibility was put forward in Refs.~\cite{mct,mctd} with the Master
Constraint  Quantization. It arises with the purpose of treating constraints that lead to quantum anomalies but allows to treat other pathological situations as we shall see.
The Master Constraint Programme tries to replace the (possibly
complicated) algebra by a much simpler Master Constraint Algebra given by the
single Master Constraint ${\bf M}$,  defined basically as a quadratic form of
the original constraints and which commutes with itself. Although this
quantization scheme fails to  detect weak Dirac observables by means of linear
conditions on them, this problem is avoided by considering second order
conditions. This approach has been successfully tested in many
situations~\cite{mctd-list,mctd-sl2r}.

Finally, a recent approach, known as the Uniform Discretizations~\cite{ud1},
has emerged in parallel following in part the lines of the Master Constraint
Quantization and attempting to improve other discrete quantization procedures 
known as consistent discretizations~\cite{cd}. In fact, this new paradigm
essentially consists in recovering the original, continuum theory from a set of
discrete theories, just like, e.g., continuum QCD can be recovered from lattice
QCD. The advantage that this new approach presents is that one starts with a
discrete version of the continuum theory that is under control, free of
drawbacks, and where a consistent quantum description is available, a priori,
since one is working off-shell within this approach, having no further constraints to be imposed. In consequence, the discrete theory contains a higher number of degrees of freedom with
respect to the continuum one, but keeping in mind that the latter can be
systematically obtained from its discrete version thanks to the existence of
certain conserved quantities (with respect to the evolution) characterizing the
continuum limit. Additionally, it succeeds in identifying both discrete and
continuum Dirac observables. That is, given a discrete constant of motion one
can identify its corresponding perennial in the continuum limit, and viceversa,
for a given perennial  in the continuum model there are in general many discrete
constants of the motion associated with. 

The main purpose of this manuscript is to confront these different quantization 
schemes in a simple but non-trivial, totally constrained system, which is
characterized by two Hamiltonians and one diffeomorphism constraints satisfying
an $sl(2,\mathbb{R})$ Lie algebra. This toy model was originally introduced in 
Ref.~\cite{mrt} testing so the Dirac approach, together with a deeper analysis 
regarding its dynamics carried out in Ref.~\cite{gambp}. Besides, a considerable
 number of publications~\cite{trunk,loukor,loukom} has appeared within the
Algebraic Quantization, and also when testing the Master Constraint
Programme~\cite{mctd-sl2r}. In both schemes, the constraints can be imposed
simultaneously at the quantum level, but the final physical Hilbert space
requires additional inputs for achieving a semiclassical limit compatible with
the  classical theory. More specifically, in Algebraic Quantization
approach~\cite{mrt,trunk,loukor,loukom} the group averaging technique cannot be
suitably applied since the symmetry group is non-amenable. Therefore, in these
cases, one has to appeal to the reality conditions of a given family of Dirac
observables in order to determine the inner product of the physical Hilbert
space. Similarly, in the Master Constraint Programme~\cite{mctd-sl2r}, the
physical Hilbert space with the correct semiclassical limit is obtained after
incorporating in the Master Constraint additional information about the Dirac
observables or an additional condition tantamount to certain specific  linear combination of constraints with non trivial coefficients dependent on the kinematical variables.

In conclusion, all this proposals incorporate at their fundamental level
information  about the observables. But in more  realistic settings, as gravity,
this requirement can defeat the completion of the quantum description,
essentially because of the difficulty that entails the identification of Dirac
observables. For just such an eventuality, it seems natural to investigate
different possibilities, like the one we are proposing in this manuscript. In
particular, we adhere to the Uniform Discretizations approach for the
quantization of a fully constrained $SL(2,\mathbb{R})$ model. The Hamiltonian
which defines the discrete (off-shell) evolution coincides in form with the
Master Constraint~\cite{mctd-sl2r}. After quantization, we study its spectrum,
paying special attention to the lowest eigenvalues, since they will
provide the best candidates for semiclassical states.  There, we identify
sectors (whose states have finite norm) that naturally induce
restrictions to the accessible  expectation values of a given subset of
observables. We consider these subspaces as the semiclassical sectors of
the theory. At this level, we follow two strategies. One consists of strictly restricting the study to one of these subspaces, which requires to modify the observable algebra quantum-mechanically, such
that the modified algebra leave invariant these sectors while
reproduces a semiclassical limit compatible with the continuum theory.
Among these subspaces, the best candidate corresponds to the lowest
possible eigenvalue, providing a similar description to the one of the Master
Constraint proposal \cite{mctd-sl2r}. The alternative strategy deals with the study of the quantum dynamics
in the context of the Uniform Discretizations. It is well known that any
quantum mechanical system evolves (non trivially) whenever there are available
several (at least two) energy states. With this in mind, we consider all the
subspaces corresponding to the lowest eigenvalues of the Hamiltonian. Among them,
we discuss about possible semiclassical sectors where the quantum
discrete dynamics would be compatible with the classical one and, moreover, with the classical constrained model, providing robustness to
the physical predictions of the Uniform Discretizations.

Hence our prescription requires that at
the end of the day the theory provides a suitable semiclassical limit, with the
observables playing a secondary role since the fundamental
structure of the final quantum theory does not depend drastically on them, but just on
the requirement of compatibility with the continuous classical version of the theory. Let us
remark that this is no longer the case for the previous  proposals since they
necessarily incorporate the observables at different levels of the construction,
in order to achieve a description compatible with the classical theory. With all
this in mind, we find  the Uniform Discretizations as the most promising
quantization scheme (among the ones considered in this manuscript) since it
provides a simple and robust arena for dealing with the problem of the quantum
dynamics of this particular toy model, but with possible applications to more
general settings such as gravity.

The manuscript is organized as follows. In Sec.~\ref{sec:hom_class} we provide
a basic description about the classical system. The quantum kinematical
framework is  introduced in Sec.~\ref{sec:kinemtics}. We provide a detailed
description of the Algebraic Quantization and the Master Constraint Programme in
Secs.~\ref{sec:AQ} and~\ref{sec:MCP}, respectively. In Sec.~\ref{sec:UD} we
present the Uniform Discretizations approach. Finally, the main conclusions can
be found in Sec.~\ref{sec:conc}. We have also included the Appendixes~\ref{app:phys-states} and \ref{appB}
with additional technical details.

\section{Classical system: Kinematics, constraints and observables}\label{sec:hom_class}

The phase space of our model is composed by four configuration variables $u_1$,
$u_2$, $v_1$ and $v_2$, and their corresponding momenta $p_i$ and $\pi_i$, with
$i=1,2$. This model is endowed with three constraints
\begin{align}\label{eq:class-const}\nonumber
H_1&=\frac{1}{2}(p_1^2+p_2^2-v_1^2-v_2^2),\quad H_2=\frac{1}{2}(\pi_1^2+\pi_2^2-u_1^2-u_2^2),\\
D&=u_1p_1+u_2p_2-v_1\pi_1-v_2\pi_2,
\end{align}
whose corresponding algebra is given by
\begin{equation}\label{eq:const-alge}
\{H_1,H_2\}=D, \quad \{H_1,D\}=-2H_1, \quad \{H_2,D\}=2H_2.
\end{equation}
The total Hamiltonian of this classical theory is
\begin{equation}\label{eq:class-hamiltonian}
H_T=NH_1+MH_2+\lambda D,
\end{equation}
with $N$, $M$ and $\lambda$ playing the role of Lagrange multipliers, i.e., they do
not correspond to dynamical variables. The equations of motion of the phase space variables can be easily  computed, yielding
\begin{align}\label{eq:class-eoms}\nonumber
&\dot u_i=Mp_i+\lambda u_i,\quad \dot v_i=M\pi_i-\lambda v_i,\\
&\dot p_i=Mu_i-\lambda p_i,\quad \dot \pi_i=Mv_i+\lambda \pi_i,
\end{align}
for $i=1,2$, and where the dot indicates time derivation. 

The classical dynamics of this model can be studied by solving this set of equations or, equivalently, by means of the parametrized observables or evolving constants \cite{mrt,evol}, as we are going to discuss in what follows. The constants of motion (or Dirac observables) of this system
\begin{align}  \label{eq:obs}
& O_{12}=u_1 p_2-p_1 u_2,\quad     O_{23}=u_2 v_1-p_2 \pi_1, \nonumber \\
& O_{13}=u_1 v_1-p_1 \pi_1, \quad  O_{24}=u_2 v_2-p_2 \pi_2, \nonumber \\
& O_{14}=u_1 v_2-p_1 \pi_2, \quad  O_{34}=\pi_1 v_2-v_1 \pi_2,
\end{align}
commute with the three constraints and reflect the global $O(2,2)$-symmetry
codified in the model. They constitute the $so(2,2)$ Lie algebra which is isomorphic to the $so(2,1)\times so(2,1)$ algebra 
\begin{align}  \label{eq:obssl}
& Q_1=\frac{1}{2}(O_{23}+O_{14}), \quad P_1=\frac{1}{2}(O_{23}-O_{14}), \nonumber \\*
& Q_2=\frac{1}{2}(-O_{13}+O_{24}), \quad P_2=\frac{1}{2}(-O_{13}-O_{24}), \nonumber \\*     
& Q_3=\frac{1}{2}(O_{12}-O_{34}), \quad P_3=\frac{1}{2}(O_{12}+O_{34}). 
\end{align}
The Poisson brackets of these observables are
\begin{equation}\label{eq:class-commut}
\{Q_i,Q_j\}=\epsilon_{ij}^{\,\,\,\,\,k}Q_k, \quad \{P_i,P_j\}=\epsilon_{ij}^{\,\,\,\,\,k}P_k, \quad \{Q_i,P_j\}=0
\end{equation}
where ${\epsilon_{ij}}^{k}=g^{lk} \epsilon_{ijl}$, with $g^{lk}$ being the
inverse of the metric $g_{lk}=\rm{diag}(1,1,-1)$. The Levi-Civita symbol
$\epsilon_{ijk}$ is totally antisymmetric with $\epsilon_{123}=1$. Besides,
repeated indexes indicates sum on them.

Together with this observables there is a reflection operator that commutes with the constraints. It is defined as
\begin{align}
R_{\epsilon_1,\epsilon_2}:&\,(u_1,u_2,v_1,v_2,p_1,p_2,\pi_1,\pi_2) \to \,(u_1,\epsilon_1u_2,v_1,\epsilon_1\epsilon_2v_2,p_1,\epsilon_1p_2,\pi_1,\epsilon_1\epsilon_2\pi_2),\label{class-reflec}
\end{align}
where $\epsilon_i=\{1,-1\}$. Its action on the observables $Q_i$ produces reflections of the type $Q_i\to -Q_i$ for $i=1,3$ (and the very same for the $P_i$), and also exchanges of the type $Q_i\longleftrightarrow P_i$ for $i=1,2,3$. 
The classical observable algebra consists of $Q_i$, $P_i$ and $R_{\epsilon_1,\epsilon_2}$, together with the commutation relations \eqref{eq:class-commut}.

Furthermore, the following identities between observables and constraints
\begin{align}\label{eq:obs-vs-const1}
&Q_1^2+Q_2^2-Q_3^2 = P_1^2+P_2^2-P_3^2=\frac{1}{4}(D^2+4H_1H_2)=:{\cal C}, \\\label{eq:obs-vs-const2}
& 4 Q_3 P_3 = (u_1^2+u_2^2)H_1-(u_1p_1+u_2p_2+v_1\pi_1+v_2\pi_2)D-(v_1^2+v_2^2)H_2,
\end{align}
will be useful from now on. We have also defined $\cal C$ as the Casimir which
commutes with all the constraints, and with the Dirac observables
(it is a combination of constraints).

The condition the three constraints vanish implies that the system has two physical degrees of freedom. The solution space has the
topology of four cones joined in the origin:
\begin{enumerate}
\item[a)] $P_3=0$ and $Q_3\in\mathbb{R}$, with $Q_1^2+Q_2^2=Q_3^2$,
\item[b)] $Q_3=0$ and $P_3\in\mathbb{R}$, with $P_1^2+P_2^2=P_3^2$,
\item[c)] $Q_3=0$ and $P_3=0$.
\end{enumerate}

Finally, and following the ideas of Ref. \cite{mrt} (see also Ref. \cite{gambp}), the kinematical variables can be expressed in terms of two independent continuous Dirac observables plus two discrete ones. Let us recall that the Dirac observables in Eq. \eqref{eq:obssl} fulfill the identities \eqref{eq:obs-vs-const1} on the constraint surface. For instance, if we define
\begin{align}
\epsilon=\frac{O_{12}}{|O_{12}|},\quad \epsilon'=\frac{O_{34}}{|O_{34}|},\quad J=|O_{12}|,\quad \tan\phi=\frac{O_{14}}{O_{13}},
\end{align}
any configuration variable can be solved in terms of these observables and the three remaining configuration variables (as well as for the momenta). More concretely, the identity
\begin{equation}
u_iv_j\epsilon^{ik}\epsilon^{jl}(u_kv_l-p_k\pi_l)=O_{12}O_{34},
\end{equation}
allows us to solve $u_1$ as
\begin{equation}
u_1=\frac{-\epsilon'u_2(v_2\cos\phi-v_1\sin\phi)+\epsilon J}{\epsilon(v_1\cos\phi+v_2\sin\phi)}.
\end{equation}

Therefore, one can define an evolving constant $U_1$ that takes the value $u_1$ when the remaining variables take the values $u_2=x$, $v_1=y$ and $v_2=z$, i.e.,
\begin{equation}
U_1=\frac{-\epsilon'x(z\cos\phi-y\sin\phi)+\epsilon J}{\epsilon(y\cos\phi+z\sin\phi)},
\end{equation}
as well for the remaining phase space variables of the model. Concluding, the gauge
invariant evolution of the model is totally captured in this description.

\section{Quantization: kinematical Hilbert space}\label{sec:kinemtics}

First of all, we will introduce the kinematical Hilbert space where
the canonical commutation relations of the phase space variables will be
defined.  In particular, we will adopt a standard Schr\"odinger representation
of square integrable functions ${\cal H}_{\rm kin}={\cal L}^2(\mathbb{R}^4)$,
and $\hbar=1$. For a given $\psi(u,v)\in {\cal H}_{\rm kin}$, the phase space
variables are promoted to the operator representation 
\begin{align}\nonumber
&\hat{p}_i\psi(u,v)=-i \partial_{u_i}\psi(u,v),\quad  \hat{\pi}_i\psi(u,v)=-i\partial_{v_i} \psi(u,v), \\\label{eq:op-reps}
& \hat{u}_i\psi(u,v) =u_i\psi(u,v), \quad
\hat{v}_i\psi(u,v) =v_i\psi(u,v).
\end{align}
Within this representation, the quantum operators corresponding to the
constraints~(\ref{eq:class-const}) are given by
\begin{align}\nonumber
&\hat H_1=-\frac{1}{2}(\partial_{u_1}^2+\partial_{u_2}^2+v_1^2+v_2^2),\\\nonumber
&\hat  H_2=-\frac{1}{2}(\partial_{v_1}^2+\partial_{v_2}^2+u_1^2+u_2^2),\\*\label{eq:quant-const}
&\hat D=-i(u_1\partial_{u_1}+u_2\partial_{u_2}-v_1\partial_{v_1}-v_2\partial_{v_2}),
\end{align}
where the factor ordering of $\hat D$ is selected such that the quantum
commutation relations agree with the classical constraint
algebra~(\ref{eq:const-alge}), that is
\begin{equation}\label{eq:q-const-alge}
[\hat H_1,\hat H_2]=i\hat D, \quad [\hat H_1,\hat D]=-2i\hat H_1, \quad [\hat H_2,\hat D]=2i\hat H_2.
\end{equation}

Regarding the observable algebra~(\ref{eq:obssl}), since their definition
involves product of commuting  phase space variables, the corresponding
operators will be free of factor ordering ambiguities. Their commutation 
relations are given by
\begin{equation}
[\hat Q_i,\hat Q_j]=i\epsilon_{ij}^{\,\,\,\,\,k}\hat Q_k, \quad [\hat P_i,\hat P_j]=i\epsilon_{ij}^{\,\,\,\,\,k}\hat P_k, \quad [\hat Q_i,\hat P_j]=0.
\end{equation}

Finally, the quantum analogues to the classical identities
(\ref{eq:obs-vs-const1}) and (\ref{eq:obs-vs-const2}) can be obtained directly
just by replacing the classical elements by their quantum versions.

\section{Algebraic Quantization and Refined Algebraic Quantization}\label{sec:AQ}

Here we will detail the Algebraic Quantization adopted in Ref.~\cite{mrt}, together with 
a brief description of the Refined Algebraic Quantization~\cite{loukor} at the end of this section.

\subsection{Algebraic Quantization}

Following the results of~\cite{mrt,loukor}, one may look for the solutions to the 
quantum constraints~(\ref{eq:quant-const}). After a transformation to polar coordinates
$u_1=u\cos\alpha$, $u_2=u\sin\alpha$, $v_1=v\cos\beta$ and $v_2=v\sin\beta$, one finds
that the corresponding solutions to the three constraints are given by
\begin{equation}\label{eq:alg-q-sol}
\Psi_{m,\epsilon}:=e^{im(\alpha+\epsilon\beta)}J_{m}(uv),
\end{equation}
where $m\in \mathbb{Z}$, $\epsilon\in\{1,-1\}$, and the functions $J_{m}(uv)$ are the
Bessel functions of first kind~\cite{abra}. With the exception of the
identity $\Psi_{0,1}=\Psi_{0,-1}$, the remaining solutions are linearly
independent~\cite{loukor}. 

Regarding the observable algebra~(\ref{eq:obssl}), let us introduce the more
convenient basis
\begin{align}
\hat Q_+ &= \frac{1}{\sqrt{2}}(\hat Q_1+i\hat Q_2),\quad \hat Q_- = \frac{1}{\sqrt{2}}(\hat Q_1-i\hat Q_2), \nonumber \\
\hat P_+ &= \frac{1}{\sqrt{2}}(\hat P_1+i\hat P_2),\quad \hat P_- = \frac{1}{\sqrt{2}}(\hat P_1-i\hat P_2),
\end{align}
with $\hat Q_3$ and $\hat P_3$ unaltered. Their commutation relations are
\begin{align}\label{eq:nobs-comm}
&\big[\hat Q_3,\hat Q_\pm \big]=\pm \hat Q_\pm, \quad  \big[\hat Q_+,\hat Q_-\big]=-\hat Q_3,\\
&\big[\hat P_3,\hat P_\pm \big]=\pm \hat P_\pm, \quad  \big[\hat P_+,\hat P_-\big]=-\hat P_3. 
\end{align}

The action of these operators on the solutions~(\ref{eq:alg-q-sol}) is 
\begin{align}
\hat Q_3 \Psi_{m,\epsilon}&= \delta_{1,\epsilon}m \Psi_{m,\epsilon}\, ,\quad \hat Q_\pm \Psi_{m,\epsilon}=\pm i\sqrt{2} \delta_{1,\epsilon}m \Psi_{m\pm 1,\epsilon}\, ,\nonumber \\
\hat P_3 \Psi_{m,\epsilon}&= \delta_{-1,\epsilon}m \Psi_{m,\epsilon}\, ,\quad \hat P_\pm \Psi_{m,\epsilon}=\pm i\sqrt{2} \delta_{-1,\epsilon}m \Psi_{m\pm 1,\epsilon}\, .
\end{align}
Now, we can identify the sectors of this solution space providing an irreducible representation of this
observable algebra, and endow them with a suitable inner product according to the adjoint relations 
\begin{eqnarray}\label{eq:re-conds}
(\hat Q_3)^{\dagger}=\hat Q_3,\;\; (\hat Q_\pm)^{\dagger}=\hat Q_\mp\,;\;\; (\hat P_3)^{\dagger}=\hat P_3,\;\; (\hat P_\pm)^{\dagger}=\hat P_\mp.
\end{eqnarray}
More specifically, the inner product is such that
\begin{align}\nonumber
&2(m\pm 1)^2(\Psi_{m,\epsilon_1=1},\Psi_{m,\epsilon_2=1})=-(\hat Q_\mp\Psi_{m\pm1,\epsilon_1=1},\hat Q_\mp\Psi_{m\pm1,\epsilon_2=1})\\
&=-(\Psi_{m\pm1,\epsilon_1=1},\hat Q_\pm\hat Q_\mp\Psi_{m\pm1,\epsilon_2=1})=2m(m\pm 1)(\Psi_{m\pm1,\epsilon_1=1},\Psi_{m\pm1,\epsilon_2=1}).
\end{align}
This condition together with the corresponding ones associated with the remaining values of $\epsilon_1$ and $\epsilon_2$ (recalling that the states $\Psi_{m,\epsilon_1}$ are orthogonal) are fulfilled if
\begin{equation}\label{eq:inner-prod}
(\Psi_{m,\epsilon_1},\Psi_{m',\epsilon_2})=a_{\epsilon_1,\epsilon_2}|m|\delta_{m,m'}\delta_{\epsilon_1,\epsilon_2},
\end{equation}
where the four constants $a_{\epsilon_1,\epsilon_2}$ do not depend on the label $m$.
Therefore, the physical Hilbert space is endowed with a basis of normalizable states $\Psi_{m,\epsilon}/\sqrt{|m|}$  that provides four irreducible representations
of the observable algebra. Each of them corresponds the completion of the subspaces
\begin{equation}
V_{\epsilon_1,\epsilon_2}:={\rm span}\{\Psi_{m,\epsilon_2}|\epsilon_1m>0\},
\end{equation}
with the inner product~(\ref{eq:inner-prod}). In the following, we will refer to each of those Hilbert spaces as ${\cal
H}_{\epsilon_1,\epsilon_2}$. The states $\Psi_{0,\epsilon}$ have zero norm. This
is one of the main handicaps of the Algebraic Quantization, since its presence
in the solution space prevents the construction of any Hilbert space. The remedy
in this case is dropping the troublesome states, something attainable  since
they are annihilated by the whole observable algebra and then they can be
decoupled from the physical Hilbert space. 

Finally, we have four different constants $a_{\epsilon_1,\epsilon_2}$ in the 
inner product~(\ref{eq:inner-prod}). Fortunately, we can take advantage of the reflection observable defined in Eq. \eqref{class-reflec} and represent it as an operator in the quantum theory. Its action on the solution space is
\begin{eqnarray} \label{eq:quant-reflec-ops}
\hat R_{\epsilon_1',\epsilon_2'}:& V_{\epsilon_1,\epsilon_2} & \mapsto  V_{\epsilon_1\epsilon_1',\epsilon_2\epsilon_2'} \nonumber \\
&\psi(u_1,u_2,v_1,v_2)&\mapsto  \psi(u_1,\epsilon_1u_2,v_1,\epsilon_1\epsilon_2v_2),
\end{eqnarray}
and it fulfills the adjoint relation $(\hat R_{\epsilon_1,\epsilon_2})^\dagger=\hat
R_{\epsilon_1,\epsilon_2}$. This last requirement restricts the previous
inner product \eqref{eq:inner-prod} and imposes that the constants $a_{\epsilon_1,\epsilon_2}$ coincide~\cite{loukor}.
One ends with a Hilbert space provided by the direct
sum of the four spaces  ${\cal H}_{\epsilon_1,\epsilon_2}$, but now equipped with the same
inner product.

Finally, the operators  $\hat Q_\pm$, $\hat Q_3$, $\hat P_\pm$ and $\hat P_3$
acting on the physical Hilbert space are
\begin{align}
&\hat Q_3 \Psi_{m,\epsilon}= \delta_{1,\epsilon}m \Psi_{m,\epsilon},\quad \hat P_3 \Psi_{m,\epsilon}= \delta_{-1,\epsilon}m \Psi_{m,\epsilon},\nonumber \\
&\hat Q_\pm \Psi_{m,\epsilon}=\pm i\sqrt{2} \delta_{1,\epsilon}\sqrt{|m(m\pm 1)|} \Psi_{m\pm 1,\epsilon},\nonumber\\
&\hat P_\pm \Psi_{m,\epsilon}=\pm i\sqrt{2} \delta_{-1,\epsilon}\sqrt{|m(m\pm 1)|} \Psi_{m\pm 1,\epsilon}.
\end{align}

\subsection{Refined Algebraic Quantization}

Regarding the Refined Algebraic Quantization of this model~\cite{loukor}, one
starts with the representation~(\ref{eq:quant-const}) and assumes that their
solutions belong to the algebraic dual $\Phi^*$ of a dense subspace $\Phi\subset
{\cal H}_{\rm kin}$. The latter is usually selected as an invariant, dense
domain of the constraints~(\ref{eq:quant-const}). The observable algebra $\cal
A_{\rm obs}$ of the model is automatically determined by these requirements, and
does not need to be included explicitly. The final step consists in introducing
the so-called rigging map~\cite{raq} between the spaces $\Phi$  and $\Phi^*$,
and  which induces an inner product in the solutions space, and then the
physical Hilbert space can be constructed out of. As was pointed out in
Ref.~\cite{loukor}, this map can be suitably defined once a convenient choice of
$\Phi$ is made. In other words, the rigging map depends on the specific choice
of test states $\Phi$, and so the observable algebra $\cal A_{\rm obs}$ and the
physical Hilbert space. In particular, the overcompleted  observable algebra
considered in the Algebraic Quantization turns out to be a subalgebra of $\cal
A_{\rm obs}$, well defined on the corresponding physical Hilbert space.
Therefore, the Refined Algebraic Quantization can be seen as a generalization of
the Algebraic Quantization.

Now, we would like to emphasize that the natural choice of test space $\Phi$
includes zero norm vectors that impede the construction of a consistent rigging
map. Then, the  solution proposed in Ref.~\cite{loukor} is to ensure that at the
end of the day the troublesome subspace is dropped by selecting a suitable test
space $\Phi$. The latter is carefully identified by means of the observable
algebra  explicitly introduced in the Algebraic Quantization, taking care that 
the observable algebra $\cal A_{\rm obs}$ remain large enough. Then, the rigging map
can be consistently constructed, and the quantum description completed.

In summary, both quantization schemes require additional information at
different levels of the construction, providing final results that
depend on it.

\section{Master Constraint Programme}\label{sec:MCP}

Now, we will continue with the quantum description within the Master Constraint
Programme \cite{mct,mctd,mctd-list,mctd-sl2r}. To this end, and for convenience, we will introduce the set of constraints
$\hat H_\pm=\hat H_1\pm \hat H_2$ with commutation relations
\begin{equation}\label{eq:const-quan-comm}
[\hat H_+,\hat H_-]=-2i\hat D, \quad [\hat H_+,\hat D]=-2i\hat H_-, \quad [\hat H_-,\hat D]=-2i\hat H_+.
\end{equation}
The Master Constraint will be defined as 
\begin{equation}
\hat {\bf M} = \frac{1}{2}(\hat H_+^2+\hat H_-^2+\hat D^2)=2\hat {\cal C}+\hat H_-^2,
\end{equation}
where we have employed the identity $4\hat {\cal C}=\hat H_+^2+\hat D^2-\hat
H_-^2$ in the previous expression.\footnote{For convenience, we have selected a Master Constraint
with a factor two with respect to the one adopted in~Ref.~\cite{mctd-sl2r}.}

In addition, we will introduce an equivalent formulation known as polarized Fock
basis (see Ref.~\cite{mctd-sl2r}). This basis is provided by the operators
\begin{equation}\label{eq:fock_A_basis}
\hat A_{\pm}:=\frac{1}{\sqrt{2}}(\hat a_1 \mp i \hat a_2), \quad \quad \hat A_{\pm}^\dagger:=\frac{1}{\sqrt{2}}(\hat a_1^\dagger \pm i \hat a_2^\dagger),
\end{equation}
and the corresponding ones for the $v$-coordinates
\begin{equation}\label{eq:fock_B_basis}
\hat B_{\pm}:=\frac{1}{\sqrt{2}}(\hat b_1 \mp i \hat b_2), \quad \quad \hat B_{\pm}^\dagger:=\frac{1}{\sqrt{2}}(\hat b_1^\dagger \pm i \hat b_2^\dagger),
\end{equation}
where 
\begin{equation}\label{eq:a-a+-variables}
\hat a_i:=\frac{1}{\sqrt{2}}(\hat u_i+i\hat p_i),\quad \hat b_i:=\frac{1}{\sqrt{2}}(\hat v_i+i\hat \pi_i),
\end{equation}
and their adjoints $\hat a_i^\dagger$ and $\hat b_i^\dagger$, are the standard
creation-annihi\-lation variables. A Fock state with respect to the annihilation
operators $\hat A_\pm$ and $\hat B_\pm$ is given by
$|k_+,k_-,k'_+,k'_-\rangle$. They are defined by means of 
\begin{equation}
|k_+,k_-,k'_+,k'_-\rangle=\frac{(\hat A_+^\dagger)^{k_+}}{\sqrt{k_+!}}\frac{(\hat A_-^\dagger)^{k_-}}{\sqrt{k_-!}}\frac{(\hat B_+^\dagger)^{k'_+}}{\sqrt{k'_+!}}\frac{(\hat B_-^\dagger)^{k'_-}}{\sqrt{k'_-!}}|0,0,0,0\rangle,
\end{equation}
where $|0,0,0,0\rangle$ is the state which is annihilated by all four
annihilation operators (in the same way that it is the vacuum state compatible
with $\hat a_i$ and $\hat b_i$ with $i=1,2$).

We will carry out a spectral decomposition of several quantities in our model. In particular, 
the observables $\hat {\cal C}$,  $\hat H_-$, $\hat Q_3$ and $\hat P_3$ can be simultaneously diagonalized 
together with the Master Constraint $\hat {\bf M}$. 

First of all, in the polarized Fock basis the observables are given by 
\begin{align}\label{eq:quan-obs}
\hat Q_\pm &=\mp\frac{i}{\sqrt{2}}(\hat A_\mp\hat B_\mp+\hat A_\pm^\dagger \hat B_\pm^\dagger), \nonumber \\*
\hat Q_3 &= \frac{1}{2}(\hat A_+^\dagger \hat A_+ -\hat A_-^\dagger \hat A_-+\hat B_+^\dagger \hat B_+ -\hat B_-^\dagger \hat B_-), \nonumber \\*
\hat P_\pm &= \mp\frac{i}{\sqrt{2}}(\hat A_\pm^\dagger \hat B_\mp^\dagger+\hat A_\mp\hat B_\pm), \nonumber \\*
\hat P_3 &= \frac{1}{2}(\hat A_+^\dagger \hat A_+ -\hat A_-^\dagger \hat A_--\hat B_+^\dagger \hat B_+ +\hat B_-^\dagger \hat B_-).
\end{align}
Since $\hat Q_3$ and $\hat P_3$ commutes with $\hat {\cal C}$ (i.e. with $\hat
{\bf M}$), we can diagonalize them simultaneously, and similarly with $\hat H_-$. This
four observables are sufficient to  identify any state of the system. 

The action of $\hat Q_3$ and $\hat P_3$ is given by
\begin{align}\label{eq:q3p3}
\hat Q_3 \, |k_+,k_-,k'_+,k'_-\rangle =q_3\,|k_+,k_-,k'_+,k'_-\rangle,   \quad q_3:=\frac{1}{2}(j-j'), \nonumber \\
\hat P_3 \, |k_+,k_-,k'_+,k'_-\rangle = p_3\,|k_+,k_-,k'_+,k'_-\rangle , \quad p_3:=\frac{1}{2}(j+j'),
\end{align}
where $j:=k_+-k_-$ and $j':= -k'_++k'_-$. Regarding the constraints $\hat H_\pm$ and $\hat D$, it is not difficult to realize that
\begin{align}
\hat H_-&=\hat A_+^\dagger \hat A_+ +\hat A_-^\dagger \hat A_- -\hat B_+^\dagger \hat B_+ -\hat B_-^\dagger \hat B_- \, ,\nonumber\\
\hat H_+&=-(\hat A_+\hat A_- + \hat A_+^\dagger \hat A_-^\dagger +\hat B_+\hat B_-+\hat B_+^\dagger \hat B_-^\dagger)\, , \nonumber \\
\hat D&=i(\hat A_+^\dagger \hat A_-^\dagger -\hat A_+ \hat A_- +\hat B_+\hat B_--\hat B_+^\dagger \hat B_-^\dagger)\, ,\label{eq:quantum-constr}
\end{align}
with the spectrum of $\hat H_-$ 
\begin{equation}
\hat H_-|k_+,k_-,k'_+,k'_-\rangle=k|k_+,k_-,k'_+,k'_-\rangle,
\end{equation}
and $k:=k_++k_--k'_+-k'_-$. Finally, we will deal with the spectral decomposition of $\hat {\bf M}$, which is
determined by the spectral properties of $\hat H_-$ and $\hat {\cal C}$ (the
Casimir). Here we will sketch the main properties that will be necessary in our
study (for a more detailed description see~\cite{mctd-sl2r} and the references therein).
Additional details can be found in \ref{app:phys-states}.

On the one hand, the spectrum of $\hat {\cal C}$ possesses both discrete  and
continuous counterparts. The discrete counterpart of $\hat {\bf M}$ is only for $k>0$ and $|j|-|j'| \geq 2$, and for
$k<0$ and $|j|-|j'|\leq 2$:
\begin{eqnarray}\label{eq:discrt-spect}
&\lambda_{\rm discr}=2 t(1-t)+ k^2, \nonumber \\
&{\rm with} \quad t=1,2,\ldots,\frac{1}{2}{\rm min}(|k|,||j|-|j'||)\,\, {\rm for\,\,even} \,\,k,  \nonumber \\
&{\rm and} \quad t=\frac{3}{2},\frac{5}{2} \ldots,\frac{1}{2}{\rm min}(|k|,||j|-|j'||)\,\, {\rm for\,\,odd} \,\,k.
\end{eqnarray}
Otherwise, the continuous part is 
\begin{eqnarray}
&\lambda_{\rm{cont}}=\frac{1}{2}+\frac{1}{2}x^2+k^2>0, \quad x \in \big[0,\infty\big) ,
\end{eqnarray}
where $x$ is independent of the particular values of $k$, $j$ and $j'$. 

The normalized eigenfunctions $|j,j'\rangle_{t,k}$ corresponding to the discrete
part of the spectrum are calculated in \ref{app:phys-states}, while the continuous
ones where determined explicitly in Ref.~\cite{mctd-sl2r}. We will use the notation
$|x,k,j,j'\rangle$ for them, which will facilitate the distinction 
between normalizable and generalized eigenstates.

As one can see, the spectrum of the Master Constraint never vanishes. Its minimum
value is in fact of the order of the square of the Planck constant (the reader
must remind that we have set it to the unity), and belongs to the continuous
part of the spectrum, corresponding to $x=0$ and $k=0$, i.e., to the eigenvalue
$\lambda_{\rm cont}=1/2$. The prescription suggested in Ref.~\cite{mctd-sl2r} 
modifies this observable by subtracting the corresponding contribution, yielding
a new Master  Constraint with a vanishing minimum eigenvalue. From now on we
will refer to this eigenspace $|x=0,k=0,j,j'\rangle$ as the physical Hilbert
space (for the Master Constraint Programme). Additionally, the restriction to
this space of solutions of the observable algebra
\begin{eqnarray}
\hat Q_3 \, |x=0,k=0,j,j'\rangle = q_3\,|x=0,k=0,j,j'\rangle,\nonumber \\
\hat P_3 \, |x=0,k=0,j,j'\rangle = p_3\,|x=0,k=0,j,j'\rangle,
\end{eqnarray}
where $q_3$ and $p_3$ are arbitrary (semi)integers --see Eqs.~(\ref{eq:q3p3})--,
indicates that the spectrum of these observables can simultaneously achieve
arbitrary large values. This is in contradiction with the classical theory,
where the condition $Q_3=0$ or $P_3=0$ (or both) must be recovered somehow.

Again, the proposal in Ref.~\cite{mctd-sl2r} consists in suitably reducing the
quantum degrees of freedom by adding to the Master Constraint the condition
$Q_3P_3=0$. Consequently, the final Master Constraint would be 
\begin{equation}
\hat {\bf M}'' = \hat {\bf M}-\frac{1}{2}\hat I + \frac{1}{2}(\hat Q_3\hat P_3)^2.
\end{equation}
The spectral decomposition of $\hat {\bf M}''$ is already known, since  $\hat
Q_3$ and $\hat P_3$ are diagonal in the basis $|x,k,j,j'\rangle$. The
restriction to the eigenspace corresponding to the minimum eigenvalue of $\hat
{\bf M}''$ yields $|x=0,k=0,j,j'\rangle$ but now with the condition~$|j|=|j'|$,
that is the requirement for a suitable semiclassical limit. Following 
Ref.~\cite{mctd-sl2r}, we will call this space of solutions $\rm SOL''$.

However, the observables~(\ref{eq:quan-obs}), while they obey the
relations (\ref{eq:obs-vs-const1}), they do not leave invariant $\rm SOL''$. Nevertheless, one can find an alternative set of observable carrying
out the relevant  physical information. In fact, the classical observables of
the type $p_1(Q_i) Q_3$ and $p_2( P_i) P_3$ with $p_1(y)$ and $p_2(y)$ being
polynomial functions of~$y$, commutes weakly with the Master Constraint ${\bf
M}''$. In consequence, any observable $p_1(Q_i) |{\rm sgn}(Q_3)|$ and $p_2(P_i)
|{\rm sgn}(P_3)|$ (with ${\rm sgn}(x) = \{1,0,-1\}$ for $x>0$, $x=0$ and $x<0$,
respectively) are also Dirac observables in the space of solutions, and then,
they leave $\rm SOL''$ invariant. 

Therefore, the basic quantum algebra will be determined by the  self-adjoint
operators $ \hat Q'_i:=|{\rm sgn}(\hat Q_3)|\hat Q_i |{\rm sgn}(\hat Q_3)|$ and
$\hat P'_i:=|{\rm sgn}(\hat P_3)| \hat P_i |{\rm sgn}(\hat P_3)|$, defined by
means of the spectral decomposition of $\hat Q_3$ and $\hat P_3$. These operators
superselect five sectors in $\rm SOL''$, one corresponding to each semiaxis of
the coordinates $q_3$ and $p_3$ satisfying the condition $q_3p_3=0$, together
with the origin $q_3=0$ and $p_3=0$. Additionally, in Ref.~\cite{mctd-sl2r} are
considered combinations of operators of the form $ |{\rm sgn}(\hat Q_3)|p_1(\hat
Q_i)|{\rm sgn}(\hat Q_3)|$ and  $ |{\rm sgn}(\hat P_3)|p_2(\hat
P_i)|{\rm sgn}(\hat P_3)|$, breaking the mentioned superselection of the
operators  $ \hat Q'_i$ and $ \hat P'_i$ regarding the coordinates $q_3$ and
$p_3$, and the resulting physical Hilbert space is given by the joint of the
five previous subspaces. 

We would like to comment that, from our point of view, the explicit inclusion of
the observable $(\hat Q_3\hat P_3)$ in  $\hat {\bf M}''$ might not be a good
prescription to be adopted for the systematic quantization of fully constrained
models owing to the difficulty of recognizing Dirac observables in more
complicated settings. Notice that even though the additional term is a
combination of constraints, there is no way of identifying its specific form without the help of the observables.

\section{Uniform Discretizations}\label{sec:UD}

\subsection{Classical description}
In this scheme \cite{ud1}, we start by considering a discrete version of a classical (continuum) theory, such that one works off-shell (but close to) with respect to the latter. On it, there is a clear notion of discrete evolution of any phase space function $F$, that is dictated by
\begin{equation}\label{eq:dscr-evol}
F_{n+1}=e^{\{\cdot,H\}}F_n:=F_n+\{F_n,H\}+\frac{1}{2!}\{\{F_n,H\},H\} +\ldots\;,
\end{equation}
with $H:=f(H_1,H_2,D)$ a well defined functional of the constraints such that
$f(x_1,x_2,x_3)$ is any non-negative function that only vanishes at the origin, it is
non-linear in the coordinates $x_1$, $x_2$ and $x_3$ and the second derivatives
satisfy
\begin{equation}
\frac{\partial^2f}{\partial x_i\partial x_j}\neq 0,\quad \forall x_i.
\end{equation}

One of the key ideas of this approach is that if one chooses initial data
such that $H=\delta^2/2$, with $\delta$ an arbitrary parameter, the constraints
will remain bounded throughout the evolution, and they will approach the
constraint surface in the limit $\delta\to 0$. On can easily realize this fact
since $H$ is itself preserved by the evolution dictated by Eq.
\eqref{eq:dscr-evol}. 

In order to follow a similar analysis like the one proposed by the  Master
Constraint approach, we will identify $H$ with the Master Constraint $\bf M$, i.e.,
\begin{equation}\label{eq:ud-hamilt}
H := 2{\cal C}+H_-^2,
\end{equation}
recalling that the Casimir ${\cal C}$ was already defined in
(\ref{eq:obs-vs-const1}). One can straightforwardly prove that $H$ satisfies all
the previous requirements as a function of the constraints. In order to analyze the classical discrete evolution, we will consider as initial data $H=\delta^2/2$. Let us start by noticing that the observables  (\ref{eq:obssl}) commutes with $H$, since
they commute with the three constraints. Consequently, their discrete evolution
is given by
\begin{equation}
(Q_i)_{n+1}=(Q_i)_{n},\quad (P_i)_{n+1}=(P_i)_{n}, \quad\forall i=1,2,3;\;n\in\mathbb{N}.
\end{equation}
They are, in consequence, constants of the motion, as well as ${\cal C}$, $H_-$
and $H$ itself. However, the constraints $H_+$ and $D$ do not commute with $H$.
In fact, one can prove that they oscillate around the surface constraint. Their
evolution is dictated by
\begin{eqnarray}\nonumber
(H_+)_{n+1} &= (H_+)_n\cos(4H_-)+D_n\sin(4H_-),\\
 D_{n+1} &= D_n\cos(4H_-)+(H_+)_n\sin(4H_-).
\end{eqnarray}
The transition matrix from an instant $n$ to $n+1$ is an $SO(2)$ rotation, of
angle $\alpha=4H_-$. From the relation of $H$ with the constraints and the initial data condition, one can easily see that $|\alpha|\leq 4\delta $. Recursively, one obtains 
\begin{eqnarray}\nonumber
&(H_+)_{n} = (H_+)_0\cos(n\alpha)-D_0\sin(n\alpha),\\ &D_{n} = D_0\cos(n\alpha)+(H_+)_0\sin(n\alpha),\label{eq:H+D-class-dynamc}
\end{eqnarray}
 where $D_0$ and $(H_+)_0$ are the initial data corresponding to each discrete
trajectory. The amplitude of the oscillations is bounded since it is given by 
\begin{equation}
0<D^2_0+(H_+)_0^2=2H-H_-^2\leq 2H=\delta^2.
\end{equation}
In fact, at any other instant $n$, the quantity $D^2_n+(H_+)_n^2$ is a constant
of motion, so the previous condition holds anytime. Clearly, in the
limit $\delta\to 0$, one recovers the continuum theory.

Let us, however, see this in more detail, by analyzing the classical discrete dynamics 
of any arbitrary phase space function $F$. In the original continuum theory, the time evolution
of any space function can be computed by means of the Poisson brackets of this phase space function with 
the classical Hamiltonian $H_T$ given in Eq. \eqref{eq:class-hamiltonian}. Therefore, we find that
\begin{equation}\label{eq:class-eom-F}
\dot F=N\{F,H_1\}+M\{F,H_2\}+\lambda \{F,D\}.
\end{equation}
Within the Uniform Discretizations, the evolution is dictated by the discrete version of this equation, i.e., Eq. \eqref{eq:dscr-evol}.
If we initially choose $H=\delta^2/2$, with $\delta$ close enough to the surface constraint, Eq. \eqref{eq:dscr-evol} simplifies since one can disregard high order contributions on $\delta$. More concretely, close to the constraint surface one has
\begin{equation}\label{eq:approx-dscr-evol}
F_{n+1}=F_n+\{F_n,H\}+O(\delta^2),
\end{equation}
with
\begin{equation}
\{F_n,H\}=H_1\{F_n,H_1\}+H_2\{F_n,H_2\}+D\{F_n,D\},
\end{equation}
and where we have omitted the label $n$ in   the constraints for simplicity. 
Now, let us write the constraints in a more convenient form
\begin{align}
&H_1=c\epsilon \cos\beta\sin\gamma, \quad H_2=c\epsilon \sin\beta\sin\gamma,\quad D=c\epsilon \cos\gamma,
\end{align}
where $c\in\mathbb{R}^+$ is an arbitrary positive parameter, $\epsilon>0$, and the angles
are $\beta\in[0,\pi)$ and $\gamma\in[0,2 \pi)$. This particular form of the
constraints (up to the points where the spherical coordinates are ill defined)
allows one to realize that $\delta=c\epsilon$. Therefore, the limit $\delta\to
0$ now amounts to $\epsilon\to 0$ for constant and finite values of $c$.

Keeping these considerations in mind, let us come back to Eq.
\eqref{eq:approx-dscr-evol} and write it in the more convenient form
\begin{align}\label{eq:disc-to-cont-class-evol}
&\frac{F_{n+1}-F_n}{\epsilon}=c \cos\beta\sin\gamma \{F_n,H_1\}+c\sin\beta\sin\gamma\{F_n,H_2\}+c \cos\gamma\{F_n,D\}+O(\epsilon).
\end{align}
If we take the limit $\epsilon\to 0$ in the previous expression (which corresponds to the limit $\delta\to 0$), the right hand
side of this equation is well defined, and the left one is identified with $\dot
F$. We then recover Eq. \eqref{eq:class-eom-F}, as we wanted to show, but with
the Lagrange multipliers determined by the initial conditions for the constraints off-shell, i.e.,
\begin{equation}
N=c \cos\beta\sin\gamma,\quad M=c\sin\beta\sin\gamma\quad \lambda=c \cos\gamma.
\end{equation}
Therefore, at a given time, the discrete dynamics reproduces in a very good approximation the
continuum classical one since any choice of the Lagrange multipliers in the continuum theory corresponds to a suitable choice of initial data off-shell for the discrete classical theory, in particular for the constraints.

\subsection{Quantization}

Now, we will study the corresponding quantization of the classical discrete
theory. In the following, we will concentrate mainly on aspects concerning the
compatibility of the discrete quantum theory with the continuum classical model,
and the advantages it presents with respect to previous approaches. We will
leave as a matter of future research the analysis of the genuine quantum discrete
dynamics involving parameterized Dirac observables also known as evolving
constants of motion \cite{gambp,evol}. For the quantum description of the model, we will adhere
to the kinematical Hilbert space explained in Sec.~\ref{sec:kinemtics}. We
represent the operators $\hat u_i$, $\hat p_i$, $\hat v_i$ and $\hat \pi_i$ with
$i=1,2$ in ${\cal H}_{\rm kin}={\cal L}^2(\mathbb{R}^4)$, as was done in
Eq.~(\ref{eq:op-reps}). Hence, the Hamiltonian~(\ref{eq:ud-hamilt}) is promoted
to the operator $\hat H$. Its spectral decomposition is the same as the one
carried out for the Master Constraint $\hat {\bf M}$. Then, the spectrum of
$\hat H$ is
\begin{eqnarray}\label{eq:Hdiscrt-spect}
&\lambda^H_{\rm discr}=2 t(1-t)+ k^2, \nonumber \\
&{\rm with} \quad t=1,2,\ldots,\frac{1}{2}{\rm min}(|k|,||j|-|j'||)\,\, {\rm for\,\,even} \,\,k,  \nonumber \\
&{\rm and} \quad t=\frac{3}{2},\frac{5}{2} \ldots,\frac{1}{2}{\rm min}(|k|,||j|-|j'||)\,\, {\rm for\,\,odd} \,\,k,
\end{eqnarray}
for the discrete counterpart, with $k>0$ and $|j|-|j'| \geq 2$ or
$k<0$ for $|j|-|j'|\leq 2$, and
\begin{eqnarray}
&\lambda^H_{\rm{cont}}=\frac{1}{2}+\frac{1}{2}x^2+k^2>0, \quad x \in \big[0,\infty\big),
\end{eqnarray}
being its continuous spectrum.

The minimum eigenvalue of $\hat H$ is provided by $k=0$ and the minimum of the
spectrum  of $\hat {\cal C}$, which is in its continuous counterpart. The
restriction to it, however, does not introduce any condition to $j$ and $j'$, as
we saw in Sec.~\ref{sec:MCP}. In consequence, the spectrum of both observables $\hat Q_3$ and
$\hat P_3$ on this space can achieve any arbitrarily large value simultaneously.

This is a fundamental aspect that force us to consider alternative
possibilities, like extending our study to other eigenspaces of $\hat H$. For instance, in its discrete spectrum, the
minimum is provided by $t=1$ and $k=\pm 2$, with no obvious restrictions for
$(|j|-|j'|)$ --see (\ref{eq:Hdiscrt-spect})--. Then, on this subspace, the
observables $\hat Q_3$ and $\hat P_3$ are not compatible with the continuum
theory, like in the subspace related to the minimum of $\lambda^H_{\rm{cont}}$.
Nevertheless, we can consider instead the whole
infrared spectrum of $\hat H$ ---i.e. those eigenvalues (non-negative real numbers) lower or of the order of $\hbar^2$--- that is compatible with, at least, certain
subalgebra of observables. 

Specifically, any state satisfying $2t<|k|<\lambda^H_{\rm discr}$
provides in fact satisfactory restrictions to the possible values of $|j|-|j'|$, 
compatible with the continuum theory (up to quantum corrections). For a given
\begin{equation}\label{eq:phys-states}
\lambda^H_{\rm discr} = 2t(1-t)+k^2,\quad {\rm with}\quad 2t<|k|<\lambda^H_{\rm discr}\;,
\end{equation}
which implies $||j|-|j'||=2t$, as one can deduce from the definition of the
discrete eigenvalues $\lambda^H_{\rm discr}$ in Eq.~(\ref{eq:Hdiscrt-spect}).
Consequently, we have two subspaces labeled by $\pm k$. Each of them, in turn,
can be split in 
\begin{equation}\label{eq:jconds}
|j|-|j'|=\pm 2t.
\end{equation}

If we recall the definition of $q_3$ and $p_3$ (the eigenvalues of $\hat Q_3$
and $\hat P_3$, respectively) given in Eq.~(\ref{eq:q3p3}), the previous
condition~(\ref{eq:jconds}) is equivalent to
\begin{equation}
|q_3+p_3|-|p_3-q_3|=\pm 2t.
\end{equation}
A simple inspection yields
\begin{equation}\label{eq:qpconds}
q_3=\pm t \quad {\rm and} \quad |p_3|\geq t,\quad {\rm or} \quad p_3=\pm t \quad {\rm and} \quad |q_3|\geq t.
\end{equation}

From now on, we will call $|q_3,p_3\rangle_{t,k}$ and $|x,k,q_3,p_3\rangle$ the
normalizable and generalized eigenstates of $\hat H$, respectively, where we
employ the labels $q_3$ and $p_3$ instead of~$j$ and~$j'$, in order to 
distinguish between our approach and the Master Constrain one. The subspace $\{|q_3,p_3\rangle_{t,k}\}$ with $q_3$ and $p_3$
fulfilling~(\ref{eq:phys-states}) --and consequently~(\ref{eq:qpconds})-- is 2-fold degenerated since $k>2t$ if $k>0$ and $k<-2t$ if $k<0$
(the specific expressions for these states can be found in~\ref{app:phys-states}). 

Among them, the states with the lowest eigenvalue of the Hamiltonian operator compatible with condition $2t<|k|<\lambda^H_{\rm discr}$ correspond to $\lambda^H_{\rm{disc}}=16$, and consequently $k=\pm 4$ and $t=1$. These states yield the best approximation to the classical theory. In this sense, this subspace is singled out from a physical point of view among the remaining ones. The first quantum description that we propose consists in restricting the study to this subspace of states. As we will see later, this proposal is similar to the one provided by the Master Constraint \cite{mctd-sl2r}. Let us also comment that any other eigenvalue $\lambda^H_{\rm{discr}}$ with the previous restriction and close to the lowest eigenvalue of $\hat H$ would give a suitable quantum description. To complete the quantization, we need to identify the observables that leave invariant each of these spaces, and particularly the one for $\lambda^H_{\rm{discr}}=16$.

\subsection{Observable algebra}

The action of the observables (\ref{eq:quan-obs}) will be easily deduced from
their commutation relations (up to a global phase), instead of a direct
calculation involving a considerable  number of algebraic manipulations. The
phase will be then straightforwardly deduced.

Let us restrict the study to the observables $\hat Q_\pm$, since the analysis
applies directly to the $\hat P_\pm$ ones. Recalling that the commutation
relations of these observables are 
\begin{equation}\label{eq:obs-comm}
\big[\hat Q_3,\hat Q_\pm \big]=\pm \hat Q_\pm, \quad  \big[\hat Q_+,\hat Q_-\big]=-\hat Q_3, 
\end{equation}
and that the Casimir operator is
\begin{equation}
\hat{\cal C}=\hat Q_+\hat Q_-+\hat Q_-\hat Q_+-\hat Q_3^2,
\end{equation}
one can easily solve, thanks to the commutation relations (\ref{eq:obs-comm}),
\begin{equation}\label{eq:pmrelat}
2\hat Q_+\hat Q_-=\hat Q_3^2-\hat Q_3+\hat{\cal C}, \quad  2\hat Q_-\hat Q_+=\hat Q_3^2+\hat Q_3+\hat{\cal C}.
\end{equation}

Having said that, and recalling that the states $|q_3,p_3\rangle_{t,k}$ are normalized
eigenfunctions of $\hat Q_3$ with eigenvalue $q_3$, from the commutation
relations (\ref{eq:obs-comm}) we deduce that  $\hat Q_\pm|q_3,p_3\rangle_{t,k}$
is either zero or proportional to $|q_3\pm 1,p_3\rangle_{t,k}$, respectively. From the
relations (\ref{eq:pmrelat}), we get
\begin{align}\nonumber
&_{t,k}\langle q_3\pm 1,p_3 |q_3\pm 1,p_3\rangle_{t,k}=\,_{t,k}\langle q_3,p_3 |(\hat Q_\pm)^\dagger \hat Q_\pm|q_3,p_3 \rangle_{t,k}=\,_{t,k}\langle q_3,p_3 |\hat Q_\mp \hat Q_\pm|q_3,p_3 \rangle_{t,k}=\\\nonumber
&\,_{t,k}\langle q_3,p_3 | \hat Q_3^2\pm\hat Q_3+\hat{\cal C}|q_3,p_3 \rangle_{t,k}= \frac{1}{2}q_3^2\pm \frac{1}{2}q_3+2t(1-t).
\end{align}
Now, let assume that 
\begin{equation}
\hat Q_\pm|q_3,p_3 \rangle_{t,k}=q_{\pm}(q_3)|q_3\pm1,p_3 \rangle_{t,k}.
\end{equation}
Hence
\begin{align}
&\hat Q_\mp \hat Q_\pm|q_3,p_3 \rangle_{t,k}=q_{\mp}(q_3\pm 1)q_{\pm}(q_3) |q_3,p_3 \rangle_{t,k}=\left[\frac{1}{2}q_3^2\pm \frac{1}{2}q_3+2t(1-t)\right] |q_3,p_3 \rangle_{t,k}\;.
\end{align}
Furthermore
\begin{align}
&q_{+}(q_3)=\,_{t,k}\langle q_3+ 1,p_3 |\hat Q_+|q_3,p_3 \rangle_{t,k}=\overline{\,_{t,k}\langle q_3,p_3 |\hat Q_-|q_3+1,p_3 \rangle_{t,k}} =\overline{q_{-}(q_3+1)}.
\end{align}
The solution to these equations is given by
\begin{align}\nonumber
&q_{\pm}(q_3)=\frac{z_\pm(q_3)}{\sqrt{2}}(q_3\pm t), \quad {\rm and } \quad z_+(q_3)z_-(q_3+1)=1,
\end{align}
with $|z_{\pm}|=1$. Consequently, the states are determined up to a global phase. 

The last step consists in determining this phase. The observables defined in Eqs.~(\ref{eq:quan-obs}), up to the global factor $i$, are a linear combination of
products (second order polynomial) of the operators $\hat A_\pm^\dagger $, $\hat
A_\pm $, $\hat B_\pm^\dagger$ and $\hat B_\pm$. Now, consider the basis elements
$|k_+,k_-,k'_+,k'_-\rangle$ of the polarized Fock quantization. The states
$|q_3,p_3\rangle_{t,k}$ are linear combinations of these basis elements, with
real coefficients. Besides, the action of the previous operators on a given
state of the polarized basis turns out to be a linear combination of the
elements of the basis, with also real coefficients. This allows us to conclude 
that, up to a global irrelevant sign, $z_\pm =\mp i$.

Finally,  in the basis $|q_3,p_3\rangle_{t,k}$, 
\begin{align}
\hat Q_+ |q_3,p_3\rangle_{t,k} &=\frac{-i}{\sqrt{2}}[q_3+t]\,|q_3+1,p_3\rangle_{t,k}  \nonumber \\*
\hat Q_-|q_3,p_3\rangle_{t,k} &= \frac{i}{\sqrt{2}}[q_3 - t]\,|q_3-1,p_3\rangle_{t,k} \nonumber \\*
\hat Q_3|q_3,p_3\rangle_{t,k}&=q_3\, |q_3,p_3\rangle_{t,k} \nonumber \\*
\hat P_+ |q_3,p_3\rangle_{t,k}&=\frac{-i}{\sqrt{2}}[p_3+t]\,|q_3,p_3+1\rangle_{t,k}\nonumber \\*
\hat P_- |q_3,p_3\rangle_{t,k}&=\frac{i}{2\sqrt{2}}[p_3-t]\,|q_3,p_3-1\rangle_{t,k} \nonumber \\*
\hat P_3|q_3,p_3\rangle_{t,k}&=p_3|q_3,p_3\rangle_{t,k} 
\end{align}

Therefore, even if one starts with an state fulfilling (\ref{eq:phys-states}) ---the one that reproduces a good semiclassical limit for $\hat Q_3$ and $\hat P_3$---, the
repeated action of the observables $\hat Q_\pm$ and $\hat P_\pm$ would turn out in a state that is not compatible with condition (\ref{eq:phys-states}) ---unless we consider on this subspace states with arbitrary large values of $k$, then losing the
semiclassical limit---.  

\subsection{Modified observable algebra}\label{sec:mod-obs-alg}

In order to overcome this drawback, we will present here a prescription  of a
modified observable algebra, based partially on the new observables introduced at
the end  of Sec.~\ref{sec:MCP}.

Let us define, appealing to the spectral theorem, the following operator
\begin{equation}
\hat t = \frac{1}{2}\hat I+\sqrt{\frac{1}{4}\hat I-\hat {\cal C}_{\rm disc}},
\end{equation}
with $\hat {\cal C}_{\rm disc}$ the restriction of the Casimir to its
discrete spectrum and $\hat I$ the identity on ${\cal H}_{\rm kin}={\cal
L}^2(\mathbb{R}^4)$. The operator $\hat t$  has a discrete spectrum that equals
the values of the parameter $t$ in Eq.~(\ref{eq:discrt-spect}). 

We will also define 
\begin{equation}\label{eq:proj-q}
\hat\varepsilon_q :=\hat I - \delta_{|\hat Q_3|,\hat t}\;.
\end{equation}
The spectrum of this operator is equal to the unity when $q_3\neq \pm t$, and zero if
$q_3=\pm t$. Similarly, we define the operator
\begin{equation}\label{eq:proj-p}
\hat\varepsilon_p :=\hat I - \delta_{|\hat P_3|,\hat t}.
\end{equation}
which are the identity when $p_3\neq \pm t$, and zero in the subspaces $p_3=\pm
t$. These operators mimic the action of the operators $|{\rm sgn}(\hat Q_3)|$
and $|{\rm sgn}(\hat P_3)|$, respectively, employed in the definitions of $\hat
Q_3'$ and $\hat P_3'$ (see the end of Sec.~\ref{sec:MCP}).

Our new modified algebra will consist in the original $\hat Q_3$ and $\hat P_3$, and the
modified ladder operators 
\begin{equation}
{\tilde Q}_{\pm}:=\hat\varepsilon_q\hat
Q_{\pm}\hat\varepsilon_q,\quad {\tilde P}_{\pm}:=\hat\varepsilon_p\hat
P_{\pm}\hat\varepsilon_p.
\end{equation}

Their action in a given space $|q_3,p_3\rangle_{t,k}$ is 
\begin{align}
&{\tilde Q}_{+}|q_3,p_3\rangle_{t,k}=(1-\delta_{|q_3|,t})(1-\delta_{|q_3+1|,t})\frac{-i}{\sqrt{2}}[q_3+t]\,|q_3+1,p_3\rangle_{t,k},\nonumber \\* 
&{\tilde Q}_{-} |q_3,p_3\rangle_{t,k}=(1-\delta_{|q_3|,t})(1-\delta_{|q_3-1|,t})\frac{i}{\sqrt{2}}[q_3-t]\,|q_3-1,p_3\rangle_{t,k},\nonumber \\*
&\hat Q_3|q_3,p_3\rangle_{t,k}=q_3\, |q_3,p_3\rangle_{t,k}, \nonumber \\*
&{\tilde P}_{+} |q_3,p_3\rangle_{t,k}=(1-\delta_{|p_3|,t})(1-\delta_{|p_3+1|,t})\frac{-i}{\sqrt{2}}[p_3+t]\,|q_3,p_3+1\rangle_{t,k},\nonumber \\*
&{\tilde P}_{-} |q_3,p_3\rangle_{t,k}=(1-\delta_{|p_3|,t})(1-\delta_{|p_3-1|,t})\frac{i}{\sqrt{2}}[p_3-t]\,|q_3,p_3-1\rangle_{t,k}, \nonumber \\*
&\hat P_3|q_3,p_3\rangle_{t,k}=p_3|q_3,p_3\rangle_{t,k}.\label{eq:quant-mod-obs}
\end{align}

From this observable algebra, we deduce that i) the four states $|q_3=\pm
t,p_3=\pm t\rangle_{t,k}$ remain invariant under the action of all the previous
observables, and ii) the subspaces 
\begin{eqnarray}\nonumber
&\{|q_3=\pm t,p_3> t\rangle_{t,k}\},\quad \{|q_3=\pm t,p_3<
-t\rangle_{t,k}\},\\& \{|q_3> t,p_3=\pm t\rangle_{t,k}\}\quad {\rm and} \quad \{|q_3<
-t,p_3=\pm t\rangle_{t,k}\},
\end{eqnarray} 
are also left invariant under this modified observable algebra. Besides, these
new observables together with the previous subspaces provide a semiclassical
limit in agreement with the Algebraic Quantization and the Master
Constraint Programme. 

Eventually, the classical reflection observables defined in Eq. \eqref{class-reflec}
can be represented as a discrete quantum operator. It preserves the sectors associated with each pair of quantum numbers $(t,k)$ but
maps each of the previous subspaces (in which these sectors can be divided) among them.\footnote{This reflection operator plays a similar role than the one of the observables $ |{\rm sgn}(\hat Q_3)|p_1(\hat
Q_i)|{\rm sgn}(\hat Q_3)|$ and  $ |{\rm sgn}(\hat P_3)|p_2(\hat
P_i)|{\rm sgn}(\hat P_3)|$ that break the superselection sectors within the Master Constraint \cite{mctd-sl2r}.}

Therefore, this quantum operator plus the modified observable algebra
\eqref{eq:quant-mod-obs}, together with the restriction to the  
sector corresponding to the lowest admissible eigenvalue
$\lambda^H_{\rm{discr}}=16$ and $t=1$ provide the final physical Hilbert space. In particular, it is the direct sum
\begin{align}
&\bigoplus_{q_3,p_3}\bigg[\bigoplus_{\epsilon_1=\pm 1}\bigoplus_{\epsilon_2=\pm1}\Big(|\epsilon_1 q_3=
1,\epsilon_2p_3= 1\rangle_{1,k}\oplus|\epsilon_1 q_3>
1,\epsilon_2p_3= 1\rangle_{1,k}\oplus|\epsilon_1 q_3=
1,\epsilon_2p_3> 1\rangle_{1,k}\Big)\bigg],\label{eq:phys-space}
\end{align}
and is 2-fold degenerated since $k=\pm 4$.

It is worth commenting that, nevertheless, the observable $\hat Q_3 \hat P_3$ is
not bounded on the physical space. This is one of the differences with the
Algebraic Quantization and the Master Constraint Programme, where the previous
quantity identically vanishes on physical solutions. Clearly, the semiclassical condition $q_3
p_3\simeq 0+O(t^2)$ for the eigenvalues of $\hat Q_3$ and  $\hat P_3$ is more restrictive than $q_3\simeq 0+O(t)$ and/or $ p_3\simeq 0+O(t)$, where the symbol $O(t^n)$ indicates contributions of the order of $t^n$ and higher. In
our proposal, the latter is satisfied while the former do not. Nevertheless, in
the limit $\hbar\to 0$, both conditions are equivalent and the continuum classical theory
is always recovered. Besides, the considerations explained before seem to be the best one can do within the Uniform Discretizations, as well as in the Master Constraint Programme by direct application, in order to achieve a suitable semiclassical description without including non-trivial contributions of the type $(\hat Q_3 \hat P_3)^2$. Concretely, the classical counterpart of this quantity fulfills the identity~\eqref{eq:obs-vs-const2}. On the one had, the left hand side of this relation is a function of (some of) the observables. The addition of this contribution involves that one needs to incorporate at least some Dirac observables of the system at the fundamental level of the approach. On the other hand, the right hand side of Eq.~\eqref{eq:obs-vs-const2} is a linear combination of constraints that involves coefficients depending on phase space. It is therefore legitimate its inclusion, for instance, within the Master Constraint approach \cite{mctd-sl2r}, since it is just a constraint. However, it is unclear how this specific condition can be inferred without the previous knowledge of some of the observables of the model.
In this context, as we have pointed out before, such type of considerations would make extremely difficult to extend this approach to more general situations like gravity.

\subsection{Discrete quantum dynamics} 

Let us now study the dynamics of this particular model at the quantum level. It can be analyzed in two different ways. One of them is by means of parametrized observables, as we mentioned at the end of Sec. \ref{sec:hom_class}, but this time restricted to the subspace $\lambda^H_{\rm{discr}}=16$ and $t=1$. However, we will leave this analysis for a future research, and concentrate in the second perspective that we mentioned at the beginning of this manuscript. It concerns the genuine quantum discrete evolution of the Uniform Discretizations \cite{ud1}. In the Heisenberg picture, the
quantum dynamics is dictated by the unitary operator
\begin{equation}
\hat U=e^{-i\hat H},
\end{equation}
(keeping in mind that we have chosen $\hbar=1$). The quantum version of Eq. \eqref{eq:dscr-evol} is given by
\begin{equation}\label{eq:quantum-dynamics}
\hat F_n=\hat U^{-1}\hat F_{n-1}\hat U=\hat U^{-n}\hat F_0\hat U^n,
\end{equation}
for any quantum observable $\hat F_0$ defined on the initial time section. From this point of view, the evolution becomes more interesting since  the Hamiltonian possesses non-vanishing 
eigenvalues. Therefore, we will not restrict the study to a subspace associated with a particular eigenvalue
of the Hamiltonian, but instead we will consider all its lowest eigenvalues. As we will see in a particular example, states peaked around the subspaces fulfilling \eqref{eq:phys-states} will provide a good semiclassical description. Within this picture, it is expected
that those observables that do not commute with the Hamiltonian will show a non-trivial
discrete dynamics.

Once a particular state $|\psi\rangle$ in the kinematical Hilbert space has been chosen, the expectation value of the quantum analog to Eq. \eqref{eq:quantum-dynamics} is given by
\begin{equation}\label{eq:exp-val-disc-qd}
\langle \hat F_{n+1}\rangle_\psi=\langle\hat U^{-1}\hat F_{n}\hat U\rangle_\psi=\langle\hat U^{-n}\hat F_{0}\hat U^n\rangle_\psi
\end{equation}
Now, since we assume that the Hamiltonian is a selfadjoint operator, its
eigenstates provide a complete basis on the kinematical Hilbert space. For
simplicity, we will denote $\lambda_m$ and $\lambda$ as the eigenvalues belonging to the discrete and continuous
parts of the spectrum of this operator, respectively. Therefore, the state  $|\psi\rangle$ can
be decomposed as
\begin{equation}
|\psi\rangle=\sum_{m}\psi_{m}|\lambda_m\rangle+\int d\lambda\psi(\lambda)|\lambda\rangle,
\end{equation}
where we have also omitted the degeneration labels of each eigenvalue for simplicity. If we
introduce this in Eq. \eqref{eq:exp-val-disc-qd}, we find that
\begin{align}\label{eq:exp-val-disc-eigen}\nonumber
&\langle \hat F_{n+1}\rangle_\psi=\sum_{m,m'}e^{-i n(\lambda_m-\lambda_{m'})}\psi_{m'}^*\psi_{m}\langle\lambda_{m'}|\hat F_0|\lambda_m\rangle+\int d\lambda d\tilde\lambda e^{-i n(\lambda-\tilde \lambda)}\psi(\tilde \lambda)^*\psi(\lambda)\langle\tilde\lambda|\hat F_0|\lambda\rangle\\
&+2\Re\left[\sum_{m}\int d\lambda e^{-i n(\lambda_m-\lambda)}\langle\lambda|\hat F_0|\lambda_m\rangle\right],
\end{align}
where the simbol $\Re$ denotes the real part. Then, the discrete quantum evolution is essentially a linear combination of oscillatory functions in the discrete time $n$ multiplied by the matrix elements of $\hat F_0$ on the basis of eigenstates of $\hat H$. If the state $|\psi\rangle$ is peaked around a given $\lambda_0$, the previous equation can be simplified
\begin{align}\label{eq:exp-val-disc-eigen1}\nonumber
&\langle \hat F_{n+1}\rangle_\psi-\langle\hat F_n\rangle_\psi=\sum_{m,m'}\left(-i (\lambda_m-\lambda_{m'})\right)\psi_{m'}^*\psi_{m}\langle\lambda_{m'}|\hat F_n|\lambda_m\rangle\\\nonumber
&+\int d\lambda d\tilde\lambda\left(-i (\lambda-\tilde \lambda)\right)\psi(\tilde \lambda)^*\psi(\lambda)\langle\tilde\lambda|\hat F_n|\lambda\rangle\\
&+2\Re\left[\sum_{m}\int d\lambda\left(-i (\lambda_m-\lambda)\right)\langle\lambda|\hat F_n|\lambda_m\rangle\right]+\ldots
\end{align}
since in the sum \eqref{eq:exp-val-disc-eigen} only those eigenvalues close to
$\lambda_0$ will not be suppressed by the wave functions. In consequence, the
differences between two eigenvalues like $(\lambda-\lambda')^a$ with $a>1$ are
supposed to be negligible with respect to the corresponding linear terms (both
$\lambda$ and $\lambda'$ must be similar to $\lambda_0$ or they will be strongly
suppressed). The dots in these expressions account for those subdominant contributions. Therefore, in this approximation, the previous equation can be written
as
\begin{equation}\label{eq:simpl-q-dscr-evol}
\langle \hat F_{n+1}\rangle_\psi-\langle\hat F_n\rangle_\psi=-i\langle [\hat F_n,\hat H]\rangle_\psi+\ldots
\end{equation}
Since the Hamiltonian is a quadratic form of the constraints, the previous expression is analogous to
\begin{align}\label{eq:simpl-q-dscr-evol-1}\nonumber
&\langle \hat F_{n+1}\rangle_\psi-\langle\hat F_n\rangle_\psi=-\frac{i}{2}\langle (\hat H_+[\hat F_n,\hat H_+]+[\hat F_n,\hat H_+]\hat H_+)\rangle_\psi-\frac{i}{2}\langle (\hat H_-[\hat F_n,\hat H_-]+[\hat F_n,\hat H_-]\hat H_-)\rangle_\psi\\
&-\frac{i}{2}\langle (\hat D[\hat F_n,\hat D]+[\hat F_n,\hat D]\hat D)\rangle_\psi+\ldots
\end{align}
Now, if the state $|\psi\rangle$ fulfills in a good approximation $\hat H_\pm|\psi\rangle\simeq h_{\pm}|\psi\rangle$ and $\hat D|\psi\rangle\simeq d |\psi\rangle$, with $h_\pm$ and $d$ some real coefficients of the order of $\sqrt{\lambda_0}$, we would get 
\begin{align}\label{eq:simpl-q-dscr-evol-2}
&\langle \hat F_{n+1}\rangle_\psi-\langle\hat F_n\rangle_\psi=-ih_-\langle [\hat F_n,\hat H_-]\rangle_\psi-ih_+\langle ([\hat F_n,\hat H_+])\rangle_\psi-id\langle ([\hat F_n,\hat D])\rangle_\psi+\ldots
\end{align}
In this circumstances, it is expected that this equation will allow us to get a
good approximation of Eq. \eqref{eq:disc-to-cont-class-evol}, and therefore, of
Eq. \eqref{eq:class-eom-F}, whenever $\lambda_0\to 0$. In the case in which this
limit cannot be strictly taken (as in our particular model), it is natural to
introduce an external energy scale which would allow one to distinguish between the discrete theory and the continuum limit. For a moment, let us recover the Planck constant $\hbar$ as well as 
$c$, the speed of light. Besides, if we compare our model with the harmonic oscillator, the frequencies and the masses involved are $\omega=1$ and $m=1$. Now, since the constraints $H_\pm$ and $D$ must have units of energy, the Hamiltonian of the Uniform Discretizations must have units of action. Therefore, we must keep in mind that our Hamiltonian is in fact normalized by  $H/K$, with $K=mc^2\omega$ a suitable 
constant (with $m=1$ and $\omega=1$) determined by the constants of our theory and with units of energy over time. Besides, the eigenvalues of $\hat H$ have units of energy square, i.e., $\lambda = \hbar^2\omega^2\tilde \lambda$ where $\tilde \lambda$ is a nonvanishing constant (and with $\omega=1$). One can easily realize that the time step of the discrete evolution will be $\Delta t=\sqrt{\tilde{\lambda}}\hbar/(mc^2)$. Whenever $\sqrt{\tilde{\lambda}}$ is of the order of the unit, which corresponds to the lowest eigenvalues of the Hamiltonian $\hat H$, the time step will be of the order of $\hbar/(mc^2)$ with $m=1$. Since it is a really small physical time, the discrete evolution for few steps will give a very good approximation of the continuous evolution.

Let us also comment that the conditions required to the previous states $|\psi\rangle$ are not enough since there could be observables, like $\hat Q_3$ and $\hat P_3$, with expectation values taking any arbitrary value simultaneously. In this case, we propose an additional requirement for the semiclassical states: they must be peaked around the subspaces restricted by condition \eqref{eq:phys-states}. With this final remark, it is expected to achieve a good semiclassical description of the model. In summary, we require to the semiclassical states to be peaked around states fulfilling Eq. \eqref{eq:phys-states} and such that the expectation values of $\hat H$ be of the order of its smallest eigenvalues.

We will now study an example
where the main aspects of the discrete quantum dynamics of the model will be
discussed. Let us consider, for instance, the evolution of the operators corresponding to the constraints of the classical theory $\hat H_+$
and $\hat D$. They are two unconstrained phase space functions, which do not commute with the Hamiltonian $\hat H$. The classical discrete dynamics of the classical analogues of
these two observables is given by Eq. \eqref{eq:H+D-class-dynamc}. These
operators, as it is shown by Eqs. \eqref{eq:kpm-action} and
\eqref{eq:const-to-kpm} of App.~\ref{appB}, have a well defined action on
every eigenstate of the Hamiltonian (see App.~\ref{appB} for
comments). For the sake of simplicity, let us consider the semiclassical state
\begin{equation}\label{eq:semiclass-state}
|\psi\rangle=\frac{1}{\sqrt{2}}(|q_3,p_3\rangle_{t(k+1)}+|q_3,p_3\rangle_{t(k-1)}),
\end{equation}
with $k-1>2t+4$. We will comment later the situation in which such inequality is not fulfilled. This state is then a 
linear combination of two different normalizable eigenfunctions of the Hamiltonian $\hat H$, as well as they fulfill the condition \eqref{eq:phys-states}, i.e., they are compatible with the classical theory. Besides, we have chosen this particular superposition because the transition amplitude of $|\psi\rangle$ with itself by means of  $\hat H_+$ and/or $\hat D$ is non-vanishing. This can be easily seen, since the expectation values of the constraints on this state are
\begin{align}
&\langle\hat H_+\rangle_\psi=\frac{\sqrt{k^2-(2t-1)^2}}{2}[(-1)^{r_-(q_3,p_3,k+1,t)}+(-1)^{r_+(q_3,p_3,k-1,t)}]\\
&\langle\hat D\rangle_\psi=i\frac{\sqrt{k^2-(2t-1)^2}}{2}[(-1)^{r_-(q_3,p_3,k+1,t)}-(-1)^{r_+(q_3,p_3,k-1,t)}],
\end{align}
with the exponents $r_\pm$ some integers depending on the quantum numbers of the eigenstates.
These expectation values can be identified with the initial data section. Besides, they are proportional to Planck constant. Let us  consider now
any arbitrary time section $n$ and the corresponding operators $\hat H_{+,n}$ and $\hat{D}_n$ 
defined by means of the unitary operator $\hat U$ and Eq. \eqref{eq:quantum-dynamics}. Their expectation values are given now by
\begin{align}
&\langle\hat H_{+,n}\rangle_\psi=\langle\hat H_{+}\rangle_\psi\cos(4kn)-\langle\hat D\rangle_\psi\sin(4kn),\\
&\langle\hat D_n\rangle_\psi=\langle\hat D\rangle_\psi\cos(4kn)+\langle\hat H_{+}\rangle_\psi\sin(4kn).
\end{align}
If we compare these expectation values with the classical evolution \eqref{eq:H+D-class-dynamc} of the constraints at different time instants, we see that both descriptions share several similarities. In both cases, the classical constraints and their corresponding expectation values in the quantum theory simply oscillate around a constant initial data, that in the latter is provided by the expectation value of the constraints on the state $|\psi\rangle$. Let us recall that the amplitude of the oscillations will be of the order of $\hbar\omega\sqrt{k^2-(2t-1)^2}$ with $\omega=1$. Besides, one can easily realize that the frequency of the oscillations corresponding to the discrete time $n$ of the classical and the quantum descriptions agree (though this is a consequence of the particular state $|\psi\rangle$ under consideration), where in the latter it will be proportional to $(\hbar\omega^2k/K)=\hbar\omega k/mc^2$, with $\omega=1$, $m=1$ and $k$ the eigenvalue of $\hat H_-$.

Let us recall that we have considered a linear combination of states in Eq. \eqref{eq:semiclass-state} such that they satisfy $k-1>2t+4$. This requirement has been adopted in order to the states $|q_3,p_3\rangle_{t(k+1)}$ and $|q_3,p_3\rangle_{t(k-1)}$ fulfill condition \eqref{eq:phys-states}. If this is not true for both states, we still would require that at least one of them belong to the subspace compatible with \eqref{eq:phys-states}. For instance, let us consider that $|q_3,p_3\rangle_{t(k+1)}$ belongs to such subspace by requiring $k=2t+1$, and any arbitrary state $|\tilde q_3,\tilde p_3\rangle_{\tilde t\tilde k}$. In order to have a non-vanishing transition amplitude of $|\psi\rangle$ with itself by means of the constraints, the states $|\tilde q_3,\tilde p_3\rangle_{\tilde t\tilde k}$ should be suitably selected. This condition restricts the possible choices to $|q_3,p_3\rangle_{t(k-1)}$, which is still a state that does not fulfill condition \eqref{eq:phys-states}. However, the expectation values of the constraints computed with $|\psi\rangle$ still give a good semiclassical description for these two particular unconstrained observables.

This simple example indicates that the existence of semiclassical sectors in the kinematical Hilbert space would be sufficient in order to describe in a good approximation the classical continuum theory. Let us remark that we have selected a particular
semiclassical state and we have shown how to construct it out of the subspace of
states fulfilling Eq. \eqref{eq:phys-states}. Although alternative  choices of
semiclassical states in agreement with the classical theory would be admissible,
the previous subspace seems to be a good starting point since it provides
natural restrictions like condition \eqref{eq:qpconds}. Besides, this
description with semiclassical states does not require the restriction to any modified
observable algebra, like the one proposed in Sec. \ref{sec:mod-obs-alg}, being
possible to work with the full algebra of observables. This modified observable algebra is 
just required whenever the study is restricted to a particular eigenspace of the Hamiltonian $\hat H$, like in the case of $\lambda_{\rm disc}=16$.

Let us add a final comment. The previous semiclassical states give a good semiclassical
description just for the constraint observables $\hat H_\pm$ and $\hat D$. In the case of more general
phase space functions, these states must be generalized. However, it is well known that the identification 
of general semiclassical states in general quantum systems is not a trivial task. We believe
that the existence of the sectors fulfilling \eqref{eq:phys-states} provides an additional ingredient that
could facilitate the identification of general semiclassical states of the theory.

\section{Conclusions}\label{sec:conc}

We have considered a totally constrained system with an $SL(2,\mathbb{R})$ gauge
group. This system is sufficiently simple and manageable while carries
difficulties that could be found in   more sophisticated, totally constrained
theories, like general relativity. We have reviewed different
approaches for the quantization of this model, with special emphasis in the
different advantages and handicaps they present. In particular, the Algebraic
Quantization (and its more sophisticated version, known as Refined Algebraic
Quantization) is able to provide a quantization where the physical Hilbert space
is constructed from a subspace of the algebraic dual of a dense set of the
kinematical Hilbert space, once it has been equipped with a suitable inner
product. The main inconvenient, within this approach, is that at the end of the
day one appeals to certain observables, requiring reality conditions in order to
pick out the physical inner product. But in general models, the
identification of such an observables could be a non-trivial task. Within the
Refined Algebraic Quantization, one applies the group averaging techniques. But
this approach requires averaging within a non-amenable group, introducing
additional difficulties in order to achieve well defined integrations. Again,
this question can be overcome by selecting a suitable family of test
states~\cite{loukor}. As we have already seen, an alternate approach that is
free of some of these drawbacks is the Master Constraint Programme. The Master
Constraint possesses a minimum, non\!\! -\!\! vanishing eigenvalue where the corresponding
infinite dimensional 
eigenspace is not entirely compatible with a suitable
semiclassical limit. Hence, the proposed solution to this problem is to include
a modified Master Constraint, which is explicitly dependent on the Dirac
observables, allowing one to  restrict the study to a particular subspace where
a suitable semiclassical theory is recovered. Therefore, one again appeals to
the Dirac observables as a fundamental ingredient in the quantum description. 
Let us also comment on the fact that one could even be tempted to consider
a reduced phase space quantization (adopting gauge fixing conditions and
avoiding so the implementation of the constraints at the quantum level).
However, in this situation, one would not have a kinematical Hilbert
space structure, and one would be far from giving a
suitable answer to the inconveniences found in more realistic situations. In
fact, it is well known that in many realistic situations gauge fixings are not
able to describe the complete constrained surface.

We suggest an alternative prescription, partially based on the Master Constraint
Programme, within the Uniform Discretizations scheme. We identify the discrete
Hamiltonian (the generator of the discrete evolution) with the original Master
Constraint ---a quadratic form in the $sl(2,\mathbb{R})$ constraints---. After
quantization,  we propose relaxing the restriction to the minimum eigenvalue
adopted in the Master Constraint, and considering instead all the subspaces
associated to the lowest eigenvalues of the Hamiltonian. There we have seen
that neither the set of generalized eigenstates nor some subsets of normalizable
eigenfunctions reproduce  by themselves a correct semiclassical limit. Nevertheless, there is a subfamily of finite norm eigenstates
carrying out an inherent cut off, the ones fulfilling $2t<|k|<\lambda^H_{\rm
discr}$, compatible with a semiclassical description when certain subalgebra of
observables is considered.  Whether this cut off is just accidental or not is
something that must be understood studying alternative systems with
non-amenable, gauge groups. At this level, we can follow two strategies.
One consists of restricting the study to any of those subspaces, in particular to the lowest admissible eigenvalue, i.e., $\lambda^H_{\rm discr}=16$, $t=1$ and $k=\pm 4$. Nevertheless,
these solution spaces are
not invariant under the whole $so(2,1)\times so(2,1)$ algebra, losing the
compatibility with the classical theory. In order to overcome this inconvenient,
we modify this observable algebra at the quantum level, in such a way they have
a well defined action and leave invariant these subspaces while reproduce a
suitable semiclassical limit. This
family of states (together with the modified observable algebra) gives the best description in agreement with the classical
theory, and is analogous to the description proposed in the Master Constraint. In this situation, the dynamics of the model can be studied by means of the so-called evolving constants (or parametrized observables) \cite{evol,mrt,gambp}. The second strategy consists in the study of the genuine quantum discrete dynamics of the Uniform Discretizations. In this case we consider all the kinematical Hilbert space and we
identify there semiclassical states. We have seen that there is a suitable family of states, those fulfilling $2t<|k|<\lambda^H_{\rm
discr}$, of which it would be possible to obtain a suitable semiclassical description of the unconstrained
model with a non-trivial quantum discrete dynamics compatible with the classical (discrete and continuous) theory. More specifically, we have studied a particular example for a couple of dynamical variables, providing a (partial but) successful semiclassical description.

Our proposal can
obviously be adopted by the Master Constraint Programme in the first situation. These two approaches,
in comparison with the Algebraic Quantization, possess a kinematical structure
well adapted to the physical one, while in the latter the physical states belong
to a larger functional space where, in particular, the state $q_3=0=p_3$ is
excluded by the quantum theory. Let us emphasize that  the Uniform
Discretizations seems to carry all the relevant information  about a suitable
semiclassical description at the quantum level, without requiring any additional
input. We understand, from a conceptual point of view, that our proposal
provides a radically different perspective with respect to the other
two quantization approaches, even more if one is interested in the application
of these quantization techniques to more general totally constrained models
like, e.g., general relativity.

Finally, let us give some final remarks about the quantum dynamics of the
system. In the continuum theory, an usual strategy has been making use of the
so-called  evolving constants~\cite{evol}. This was the original point of view
adopted in Ref.~\cite{mrt}, where the  reality conditions required to the
quantum evolving constants considerably restrict the possible choices of these
quantum observables~\cite{gambp}. The Uniform Discretizations, as well as
the Master Constraint proposal of Ref. \cite{mctd-sl2r}, admits a description in
these lines, as we have already mentioned. In particular, it would be
interesting to compare them with the continuous quantization provided in Ref.
\cite{gambp}. Additionally, within the Uniform Discretizations, the freedom that we have
introduced by considering arbitrary states but, obviously, giving expectation values of $\hat H$ of the order of its lowest eigenvalues, turns out
into a non-trivial, discrete quantum dynamics, where the system can evolve since
there are many ``energy'' states available. 
From this point of view, one can also study the relational dynamics 
analyzing the conditional probabilities~\cite{ud1,cond-prob} without selecting
any particular variable as a time parameter and its consequent
treatment as a classical variable. Here, one considers the probability that a
given observable have a particular value when we make a measurement on another
one. This point of view seems to be more natural since all the variables in the kinematical space of the system can be treated
 quantum-mechanically. Then, it would be interesting to check under which approximations this relational dynamics coincides with the one resulting from the use of the evolving constants technique~\cite{mrt,gambp}. We will study all these aspects in
the future.

\section*{Appendices}
 
\appendix

\section{Physical states: normalizable solutions}\label{app:phys-states}

In this Appendix, we will describe the spectral resolution of $\hat H$ adopting
the  treatment of Ref.~\cite{mctd-sl2r}. Essentially, one starts with a 
representation of the positive and negative discrete series of
$sl(2,\mathbb{R})$. Each representation is associated with the corresponding
Hilbert spaces of holomorphic and anti-holomorphic functions on the open unit
disc in $\mathbb{C}$, respectively, endowed with the scalar product, in both
cases,
\begin{equation}
 \langle f,h\rangle_l
=\frac{l-1}{\pi}\int_{D} f(z) \overline{h(z)} (1-|z|^2)^{l-2} dx\,dy 
\end{equation}
where $D$ is the unit disc and $dx dy$ is the Lebesgue measure on $\mathbb{C}$.
If $l=1$ one simply considers the limit $l\to 1$ in the previous expression.

For the positive series, an orthonormal basis is given by
\begin{eqnarray}
f^{l}_n:=\Big[\mu_{l}(n)\Big]^{-\frac{1}{2}}z^{n}\quad  (n \in \mathbb{N})\quad  { \rm{with}}\quad \mu_{l}(n)=\frac{\Gamma(n+1)\Gamma(l)}{\Gamma(l+n)} ,
\end{eqnarray}
while for the negative series, the corresponding basis is given by the complex
conjugated of~$f^{l}_n$. There exists also a unitary map between the polarized
basis $\{|k_+,k_-,k'_+,k'_-\rangle\}$ and the basis  provided by $f^{l}_n\otimes
(f^{l'}_{n'})^*$, given by
\begin{align} \label{unita}
U: f^{|j|+1}_n\otimes \left(f^{|j'|+1}_{n'}\right)^*  \mapsto & (-1)^{n'}|k_+,k_-,k'_+,k'_-\rangle \quad \rm{where} \nonumber \\*
& 2 n=k_++k_--|j|\, , \quad j=k_+-k_- \, , \nonumber \\*
& 2 n'=k'_++k'_--|j'|\, , \quad j'=-k'_++k'_- .
\end{align}

In this representation the Master Constraint is a differential operator~\cite{mctd-sl2r}, whose
eigenfunctions are  of the form 
\begin{equation}\label{eq:solu1}
f_{k,j,j'}(z_1,\overline{z}_2,t)=f_{k,j,j'}(z_1\overline{z}_2,t)\,z_1^{\frac{1}{2}(k-|j|+|j'|)},
\end{equation}  
where the solutions that are regular at $z=0$, with $z:=z_1\overline{z}_2$, are
\begin{align} \label{appendixsol1}\nonumber
&f_{k,j,j'}(z,t)=(1-z)^{1-t-\frac{1}{2}(|j|+|j'|+2)}\\*
&\times F\Big(1-t+\frac{1}{2}(-|j|+|j'|),1-t+\frac{1}{2}k,1+\frac{1}{2}(k-|j|+|j'|);z\Big),
\end{align}
for $k-|j|+|j'|\geq 0$, and 
\begin{align} \label{appendixsol2}\nonumber
&f_{k,j,j'}(z,t)=(1-z)^{1-t-\frac{1}{2}(|j|+|j'|+2)} z^{\frac{1}{2}(-k+|j|-|j'|)}  \\* &\times   F\Big(1-t-\frac{1}{2}k,1-t+\frac{1}{2}(|j|-|j'|),1+\frac{1}{2}(-k+|j|-|j'|);z\Big), 
\end{align}
for $k-|j|+|j'|\leq 0$, being $t=\frac{1}{2}(1+\sqrt{1-\lambda+2k^2}),\,
\rm{Re}(t) \geq \frac{1}{2}$ and $F(a,b,c;z)$ the hypergeometric function~\cite{abra}.

Finally, we can use the map $U$ in (\ref{unita}) to transfer these results to
the original kinematical Hilbert space ${\cal
L}^2(\mathbb{R}^4)$. To this
end we rewrite (\ref{eq:solu1}) into a power series in $z_1$ and $\overline{z}_2$
using the definition of the hypergeometric function 

\begin{equation}
F(a,b,c\,;z)=\frac{\Gamma(c)}{\Gamma(a)\Gamma(b)}\sum_{n=0}\frac{\Gamma(a+n)\Gamma(b+n)}{\Gamma(c+n)\Gamma(1+n)}z^n,
 \end{equation} 
and 
\begin{equation}\label{eq:1-zexpan}
 (1-z)^{1-d}=\sum_{n=0}\frac{\Gamma(d+n-1)}{\Gamma(d-1)\Gamma(n+1)}z^n. 
\end{equation}

For
$k-|j|+|j'|\geq 0$ we obtain
\begin{align}  \label{sol1a}
&f(t;k,j,j')=U\Big(f_{k,j,j'}(z_1,\overline{z}_2,t)\Big)=\sum_{m=0} a_m \,|k_+(m),k_-(m),k'_+(m),k'_-(m)\rangle,
\end{align}
where
\begin{align}  
k_+(m)&=m+\frac{1}{2}(k+j+|j'|),\quad   k_-(m)=m+\frac{1}{2}(k-j+|j'|), \nonumber \\*
k'_+(m)&=m+\frac{1}{2}(|j'|-j'),\quad  k'_-(m)=m+\frac{1}{2}(|j'|+j'),
\end{align}
and 
\begin{align} \label{am}
&a_m =(-1)^m \left[\mu_{(|j|+1)}\Big(m+\frac{1}{2}(k-|j|+|j'|) \Big)\right]^{\frac{1}{2}} \left[\mu_{(|j'|+1)}(m )\right]^{\frac{1}{2}}    \,\, \nonumber\\*
& \times \frac{\Gamma\Big(1+\frac{1}{2}(k-|j|+|j'|) \Big)}{\Gamma\Big(1-t+\frac{1}{2}(-|j|+|j'|) \Big)\Gamma\Big(1-t+\frac{1}{2}k \Big)} \,\,\nonumber \\*
& \times\,\sum_{l=0}^m \frac{\Gamma\Big(1-t+\frac{1}{2}(-|j|+|j'|)+l \Big) \Gamma\Big(1-t+\frac{1}{2}k+l \Big)}{\Gamma\Big(1+\frac{1}{2}(k-|j|+|j'|)+l \Big)\Gamma\Big(1+l \Big)}\\*
&\times \frac{\Gamma\Big(t+\frac{1}{2}(|j|+|j'|) +(m-l)\Big)}{\Gamma\Big(m-l+1\Big)\Gamma\Big(t+\frac{1}{2}(|j|+|j'|) \Big)} .\nonumber
\end{align}
It is worth comment that replacing $k$ with $-k$, switching $|j|$ and $|j'|$ and
multiplying with $(-1)^{\frac{1}{2}(-k+|j|-|j'|)}$, we obtain the coefficient
$a_m$ for the solution corresponding to~$k-|j|+|j'|\leq 0$.

\subsection{Normalizable eigenfunctions of $\hat H$:}

Let us focus on the normalizable eigenfunctions of $\hat H$ fulfilling the condition 
$2t<|k|<\lambda^H_{\rm discr}$. More precisely, we will start with those
such that $k-|j|+|j'|\geq 0$. Since $|j|-|j'|=\pm 2t$ and $|k|>2t$, i.e. $k\pm
2t>0$, the only possibility is that of $k>0$. If we substitute this
in~(\ref{appendixsol1}) we get
\begin{equation}
f_{\pm}(z)=(1-z)^{-t\mp t-|j'|}F\left(1-t\mp t,1-t+\frac{1}{2}k,1\mp t+\frac{1}{2}k;z\right),
\end{equation}
where
\begin{align}\nonumber
&f_{+}(z)=(1-z)^{-2t-|j'|}F\left(1-2t,1-t+\frac{1}{2}k,1- t+\frac{1}{2}k;z\right)\\*
&=(1-z)^{-|j'|-1}=\sum_{l=0}\frac{\Gamma\left(|j'|+l+1\right)}{\Gamma\left(|j'|+1\right)\Gamma\left(l+1\right)}z^{l},
\end{align}
and 
\begin{align}\nonumber
&f_{-}(z)=(1-z)^{-|j'|}F\left(1,1-t+\frac{1}{2}k,1+ t+\frac{1}{2}k;z\right)
\\*
&=\frac{\Gamma\left(1+t+\frac{1}{2}k\right)}{\Gamma\left(1-t+\frac{1}{2}k\right)}\sum_{n,l=0}\frac{\Gamma\left(1-t+\frac{1}{2}k+n\right)}{\Gamma\left(1+t+\frac{1}{2}k+n\right)}\frac{\Gamma\left(|j'|+l\right)}{\Gamma\left(|j'|\right)\Gamma\left(l+1\right)}z^{n+l}.
\end{align}

In each case, the corresponding eigenfunction (\ref{eq:solu1}) is
\begin{align}
f_{+}(z_1,\bar{z}_2)&=f_{+}(z_1\bar{z}_2)z_1^{\frac{1}{2}k-t}=\sum_{l=0}\frac{\Gamma\left(|j'|+l+1\right)}{\Gamma\left(|j'|+1\right)\Gamma\left(l+1\right)}{\bar z}_2^{l}\,z_1^{l-t+\frac{1}{2}k},
\end{align}
and 
\begin{align}\nonumber
&f_{-}(z_1,\bar{z}_2)=f_{-}(z_1\bar{z}_2)z_1^{\frac{1}{2}k+t}=\frac{\Gamma\left(1+t+\frac{1}{2}k\right)}{\Gamma\left(1-t+\frac{1}{2}k\right)}\\*
&\times\sum_{n,l=0}\frac{\Gamma\left(1-t+\frac{1}{2}k+n\right)}{\Gamma\left(1+t+\frac{1}{2}k+n\right)}\frac{\Gamma\left(|j'|+l\right)}{\Gamma\left(|j'|\right)\Gamma\left(l+1\right)}\bar{z}_2^{n+l}\,z_1^{n+l+t+\frac{1}{2}k}.
\end{align}

Let us now consider the case in which  $k-|j|+|j'|\leq 0$. Again, $|j|-|j'|=\pm 2t$ and $|k|>2t$. Since $k\pm 2t<0$, we conclude that $k<0$. Implementing all this in~(\ref{appendixsol2})
\begin{align}\nonumber
&f_{\pm}(z)=(1-z)^{-t\mp t-|j'|}z^{\pm t-\frac{1}{2}k} F\left(1-t-\frac{1}{2}k,1-t\mp t,1\mp t-\frac{1}{2}k;z\right)\\*
&=(1-z)^{-t\mp t-|j'|}z^{\pm t+\frac{1}{2}|k|}F\left(1-t+\frac{1}{2}|k|,1-t\mp t,1\mp t+\frac{1}{2}|k|;z\right),
\end{align}
or more specifically 
\begin{align}\nonumber
&f_{+}(z)=(1-z)^{-2t-|j'|}z^{t+\frac{1}{2}|k|}F\left(1-t+\frac{1}{2}|k|,1,1+ t+\frac{1}{2}|k|;z\right)
\\
&=\frac{\Gamma\left(1+t+\frac{1}{2}|k|\right)}{\Gamma\left(1-t+\frac{1}{2}|k|\right)}\sum_{n,l=0}\frac{\Gamma\left(1-t+\frac{1}{2}|k|+n\right)}{\Gamma\left(1+t+\frac{1}{2}|k|+n\right)}\frac{\Gamma\left(2t +|j'|+l\right)}{\Gamma\left(2t+|j'|\right)\Gamma\left(l+1\right)} z^{n+l+t+\frac{1}{2}|k|},
\end{align}
and
\begin{align}\nonumber
&f_{-}(z)=(1-z)^{-|j'|}z^{-t+\frac{1}{2}|k|}F\left(1-t+\frac{1}{2}|k|,1-2t,1- t+\frac{1}{2}|k|;z\right)\\*
&=(1-z)^{2t-|j'|-1}z^{-t+\frac{1}{2}|k|}=\sum_{l=0}\frac{\Gamma\left(|j'|-2t+l+1\right)}{\Gamma\left(|j'|-2t+1\right)\Gamma\left(l+1\right)}z^{l+\frac{1}{2}|k|-t}.
\end{align}

The corresponding eigenfunctions --see (\ref{eq:solu1})-- are
\begin{align}\nonumber
&f_{+}(z_1,\bar{z}_2)=f_{+}(z_1\bar{z}_2)z_1^{-\frac{1}{2}|k|-t}=\frac{\Gamma\left(1+t+\frac{1}{2}|k|\right)}{\Gamma\left(1-t+\frac{1}{2}|k|\right)}\\*
&\times \sum_{n,l=0}\frac{\Gamma\left(1-t+\frac{1}{2}|k|+n\right)}{\Gamma\left(1+t+\frac{1}{2}|k|+n\right)}\frac{\Gamma\left(2t+|j'|+l\right)}{\Gamma\left(2t+|j'|\right)\Gamma\left(l+1\right)}z_1^{n+l}\,\bar{z}_2^{n+l+t+\frac{1}{2}|k|},
\end{align}
and
\begin{align}
&f_{-}(z_1,\bar{z}_2)=f_{-}(z_1\bar{z}_2)z_1^{-\frac{1}{2}|k|+t}=\sum_{l=0}\frac{\Gamma\left(|j'|-2t+l+1\right)}{\Gamma\left(|j'|-2t+1\right)\Gamma\left(l+1\right)}z_1^{l}\,{\bar z}_2^{l-t+\frac{1}{2}|k|}.
\end{align}

The last step in our calculations consists of applying the unitary
transformation~(\ref{unita}) to the previous functions, and normalize them.
The resulting eigenfunctions now read
\begin{equation}
f_\pm(t,k,j,j') = \sum_{m=0} a_{\pm,m} |k_+(m),k_-(m),k'_+(m),k'_-(m)\rangle,
\end{equation}
with the coefficients $a_m$ given by
\begin{enumerate}
\item[a)] $k-|j|+|j'|\geq 0$ and $|j|-|j'|=2t$,
\begin{align}
&a_{+,m}=(-1)^{m}\left[\mu_{(|j'|+2t+1)}\left(m-t+\frac{1}{2}k\right)\right]^{\frac{1}{2}} \left[\mu_{(|j'|+1)}(m )\right]^{\frac{1}{2}}\frac{\Gamma\left(|j'|+m+1\right)}{\Gamma\left(|j'|+1\right)\Gamma\left(m+1\right)}.
\end{align}

\item[b)] $k-|j|+|j'|\geq 0$ and $|j|-|j'|=-2t$,
\begin{align}
&a_{-,m}=(-1)^m\left[\mu_{(|j'|-2t+1)}\left(m+t+\frac{1}{2}k\right)\right]^{\frac{1}{2}} \left[\mu_{(|j'|+1)}(m )\right]^{\frac{1}{2}} \\*\nonumber
&\times\frac{\Gamma\left(1+t+\frac{1}{2}k\right)}{\Gamma\left(1-t+\frac{1}{2}k\right)}\sum_{l=0}^m\frac{\Gamma\left(1-t+\frac{1}{2}k+l\right)}{\Gamma\left(1+t+\frac{1}{2}k+l\right)}\frac{\Gamma\left(|j'|+m-l\right)}{\Gamma\left(|j'|\right)\Gamma\left(m-l+1\right)}.
\end{align}

\item[c)] $k-|j|+|j'|\leq 0$ and $|j|-|j'|=2t$,
\begin{align}
&a_{+,m}=(-1)^m\left[\mu_{(|j'|+2t+1)}\left(m\right)\right]^{\frac{1}{2}} \left[\mu_{(|j'|+1)}(m+t+\frac{1}{2}|k| )\right]^{\frac{1}{2}} \\*\nonumber
&\times\frac{\Gamma\left(1+t+\frac{1}{2}|k|\right)}{\Gamma\left(1-t+\frac{1}{2}|k|\right)}\sum_{l=0}^{m}\frac{\Gamma\left(1-t+\frac{1}{2}|k|+l\right)}{\Gamma\left(1+t+\frac{1}{2}|k|+l\right)}\frac{\Gamma\left(2t+|j'|+m-l\right)}{\Gamma\left(2t+|j'|\right)\Gamma\left(m-l+1\right)}.
\end{align}

\item[d)] $k-|j|+|j'|\leq 0$ and $|j|-|j'|=-2t$,
\begin{align}
&a_{-,m}=(-1)^m\left[\mu_{(|j'|-2t+1)}\left(m\right)\right]^{\frac{1}{2}} \left[\mu_{(|j'|+1)}(m-t+\frac{1}{2}|k| )\right]^{\frac{1}{2}}\\*\nonumber
& \times\frac{\Gamma\left(|j'|-2t+m+1\right)}{\Gamma\left(|j'|-2t+1\right)\Gamma\left(m+1\right)}.
\end{align}
\end{enumerate}

The normalized eigenfunctions are finally defined as
\begin{equation}\label{eq:eig-coeff}
|j,j'\rangle_{t,k}:= \sum_{m=0}\tilde a_m |k_+(m),k_-(m),k'_+(m),k'_-(m)\rangle,
\end{equation}
with
\begin{equation}
\tilde a_m =a_m\left(\sum_{l=0}|a_l|^2\right)^{-1}.
\end{equation}

The states belonging to the infrarred spectrum of $\hat H$ solve the three
constraints $\hat H_\pm$ and $\hat D$ when quantum corrections of the Planck
order are neglected.

\subsection{Algebraic Quantization and physical states:}

Given the solutions (\ref{eq:solu1}) to the Master Constraint, one can ask which
is the relation between the states annihilated by $\hat M$ and the solutions~(\ref{eq:alg-q-sol}) found
within the Algebraic Quantization approach.

They can be easily computed by means of (\ref{eq:solu1}) just solving the
equation $\hat M \Psi=0$. In this case, we set $t=1$ and $k=0$ in
(\ref{eq:solu1}). After applying the map~(\ref{unita}), the resulting solutions
are
\begin{align}
\!\!f(t=1;k=0,j=m,j'=\epsilon m)= \sum_{l=0} (-1)^l |k_+(l),k_-(l),k'_+(l),k'_-(l)\rangle,
\end{align}
where $f(t=1;k=0,j=m,j'=\epsilon m)=\Psi_{m,\epsilon}$. These states solve
simultaneously the three constraints $\hat H_\pm$ and $\hat D$. Nevertheless,
they do not belong to ${\cal H}_{\rm kin}$. Hence additional considerations
are necessary in order to endow them with Hilbert space structure.

\section{Constraint observable algebra}\label{appB}

In this Appendix we will study several properties of the quantum constraints
$\hat H_+$ and $\hat D$ defined in Eq. \eqref{eq:quantum-constr}. In particular,
we are interested in the determination of their action on the eigenfunctions of $\hat H$. This space of states \{$|q_3,p_3\rangle_{tk}$\} or \{$|x,k,q_3,p_3\rangle$\} is characterized by the eigenvalue of
the Casimir $\hat{\cal C}$ which are labeled by $t$ or $x$, depending if it corresponds to the discrete or the continuous spectrum; $k$, which is the eigenvalue of the constraint $\hat H_-$;
and $\lambda_H$, the eigenvalue of the Hamiltonian $\hat H$. For simplicity, we will restrict the study to \{$|q_3,p_3\rangle_{tk}$\}, but the very same conclusions are also valid for \{$|x,k,q_3,p_3\rangle$\}.

Since the quantum constraints fulfill the commutation relations \eqref{eq:const-quan-comm}, 
it seems natural to introduce the ladder operators 
\begin{equation}\label{eq:ladder-op}
\hat K_\pm = \hat H_+\pm i\hat D.
\end{equation}
Their commutators with the constraint $\hat H_-$ can be straightforwardly deduced by means
of the commutation relations  \eqref{eq:const-quan-comm}, yielding
\begin{equation}
[\hat H_-,\hat K_\pm]=\pm 2\hat K_\pm.
\end{equation}
Now, using these commutators, one can easily conclude that the operators $\hat K_\pm$ acting on states of the form $|q_3,p_3\rangle_{tk}$ shift the label $k$ in two units, that is, 
\begin{equation}
H_-(\hat K_\pm|q_3,p_3\rangle_{tk})=(k\pm 2)\hat K_\pm|q_3,p_3\rangle_{tk}.
\end{equation}
Moreover, the square of the norms of $\hat K_\pm|q_3,p_3\rangle_{tk}$ fulfill
\begin{equation}
||\hat K_\pm|q_3,p_3\rangle_{tk}||^2=(2\lambda_h-k^2\pm 2k)||\;|q_3,p_3\rangle_{tk}||^2.
\end{equation}
This result, together with the relations \eqref{eq:quantum-constr} and the reality of the coefficients
$\tilde a_{m}$ in Eq.~\eqref{eq:eig-coeff} for the normalized eigenfunctions $|q_3,p_3\rangle_{tk}$,
allow us to conclude that the action of the operators $\hat K_\pm$ is given by
\begin{equation}\label{eq:kpm-action}
\hat K_\pm|q_3,p_3\rangle_{tk}=(-1)^{r_\pm(q_3,p_3,k,t)}\sqrt{2\lambda_h-k^2\pm 2k}|q_3,p_3\rangle_{t(k\pm 2 )},
\end{equation}
with $r_\pm(q_3,p_3,k,t)$ some integers that can depend on the corresponding
quantum numbers, such that $r_\pm(q_3,p_3,k,t)+r_\mp(q_3,p_3,k\pm 2,t)$
are even integers. It is worth commenting that the action of $\hat K_\pm$ on the
eigenstates \{$|q_3,p_3\rangle_{tk}$\} with $k=\pm 2t$, respectively, is by annihilation. For the generalized eigenfunctions \{$|x,k,q_3,p_3\rangle$\}, it only happens for $k=\pm 1$ and $x=0$.

Now, a straightforward calculation allows us to conclude that
\begin{equation}\label{eq:const-to-kpm}
\hat H_+=\frac{1}{2}(\hat K_++\hat K_-),\quad \hat D = \frac{i}{2}(\hat K_--\hat K_+).
\end{equation}
Therefore, the action of the constraints on the solution space provided by
condition \eqref{eq:phys-states} mixes states with labels $k\pm 2$.

\section*{Acknowledgments} 

This work was supported in part by PEDECIBA (Uruguay) and the Spanish MEC
Project FIS2011-30145-C03-02. The authors acknowledge to G. A. Mena Marug\'an
for helpful discussions. The final publication is available at Springer via 
http://dx.doi.org/10.1007/s10714-014-1768-1.

\end{document}